\documentclass[10pt,a4paper]{article}
\usepackage[utf8]{inputenc}
\usepackage[english]{babel}
\usepackage[title]{appendix}
\usepackage{graphpap,amsmath,amsfonts,amssymb,epsfig,makeidx,graphicx,textcomp,cite,citeall,authblk,subfloat,float,color} 
\usepackage[left=1.5cm, right=1.5cm, top=1.5cm, bottom=1.5cm]{geometry}

\author[1]{{Soham Chandra} \thanks{E-mail addresses: soham.rs@presiuniv.ac.in ; sohamc07@gmail.com}}

\affil[1]{\textit{\normalsize{Department of Physics, Presidency University, 86/1 College Street, Kolkata -700 073, India}}}

\vspace{20pt}

\title{\textbf{Effect of a Gaussian random external magnetic field with spatiotemporal variation on compensation in Ising spin-1/2 trilayered square ferrimagnets}}

\date{} 
\begin{document}
	\maketitle	
	\begin{abstract}
		In this work, an extensive Metropolis Monte Carlo simulation is performed to investigate the steady-state magnetic and thermodynamic behaviour of a trilayered spin-1/2 Ising ferrimagnet with square monolayers, driven by external Gaussian random magnetic field with certain spatio-temporal variations. Such \textit{thin} ferrimagnetic systems exhibit compensation phenomenon and thus are potentially interesting candidates for several technological applications. Here, two distinct theoretical atoms, A and B make up the \textit{ABA} and \textit{AAB} types of configurations in which the like atoms (A-A and B-B) ferromagnetically interact and the unlike atoms (A-B) interact antiferromagnetically. Depending upon the strength of the spatio-temporally varying Gaussian random field, the compensation and critical points shift and steady-state magnetic behaviours change between the different distinct types of ferrimagnetic behaviours. The compensation phenomenon even vanishes after crossing a finite threshold of the standard deviation of the magnetic field for particular choices of the other controlling parameters. Consequently, in the Hamiltonian parameter space of both configurations, islands of ferrimagnetic phase without compensation appear within the phase area with compensation of field-free case. The areas of such islands grow with an increasing standard deviation of the external field, $\sigma$, obeying the scaling relation: $f(\sigma,A(\sigma))=\sigma^{-b}A(\sigma)$ with $b_{ABA}=1.913\pm 0.137$ and $b_{AAB}=1.625\pm 0.066$ . These values of exponents match within the statistical interval with those obtained with the uniform random magnetic field.
	\end{abstract}

%%%%%%%%%%%%%%%%%%%%%%%%%%%%%%%%%%%%%%%%%%%%%%
\newpage
\twocolumn
%%%%%%%%%%%%%%%%%%%%%%%%%%%%%%%%%%%%%%%%%%%%%%
\section{Introduction}
\label{sec_intro}
%%%%%%%%%%%%%%%%%%%%%%%%%%%%%%%%%%%%%%%%%%%%%%
In the last few decades, research on layered magnetic superlattices has shown that they can have low density, transparency and mechanical strength, which find potential applications in magnetic recording, information storage and magneto-resistive sensors \cite{Barbic}. Amongst them, few-layered ferrimagnetic materials are often found to have physical properties very different from the bulk. Though ferrimagnetism was discovered in 1948 \cite{Cullity}, the experimental interest in ferrimagnetism has grown up rapidly with the discovery of thin film growth techniques, like, metalorganic chemical vapour deposition (MOCVD) \cite{Stringfellow}, molecular-beam epitaxy (MBE) \cite{Herman}, pulsed laser deposition (PLD) \cite{Singh}, and atomic layer deposition (ALD) \cite{George}. Such experimental advancements have made the growth of bilayered \cite{Stier}, trilayered \cite{Leiner}, and multilayered \cite{Sankowski,Pradhan,Maitra} systems with desired characteristics a reality. Expectedly, theoretical and computational studies of layered magnets have also gained momentum. For a multilayered ferrimagnet, magnetizations of each of the monolayers may evolve differently with temperature. Combination of such different magnetic behaviours in specific cases, exhibit \textit{compensation}. The Compensation point for layered magnets is that specific temperature, lower than the critical temperature, where the total magnetization of the system vanishes but individual layers remain magnetically ordered \cite{Cullity}.\\

The temperature dependence of total magnetization of layered magnets with antiferromagnetic interlayer coupling, exhibiting a ferrimagnetic ground state, may show magnetic compensation. Compensation is not related to the criticality of the system but the magnetic coercivity shows singularity at the compensation point \cite{Connell,Ostorero} for some ferrimagnetic materials. Strong temperature dependence of the coercive field around the compensation point and compensation point about the room temperature, make such ferrimagnets useful for thermomagnetic recording \cite{Connell}. At the compensation point, a small driving field can reverse the
sign of magnetization. That is why, the Magnetocaloric Effect in the vicinity of compensation temperature is studied in \cite{Ma}. So the control and manipulation of the compensation phenomenon becomes an important topic of research from the point of view of theoreticians and experimentalists. A few related examples, in this direction, follow. In \cite{Mathoniere}, it has been shown that, polycrystalline molecular magnets, for example, $N(n-C_{m}H_{2m+1})_{4} Fe^{II} Fe^{III} (C_{2}O_{4})_{3} [m=3-5]$ have compensation temperatures near $30$ K depending on the type of cation $A^{+}$ . This particular kind of system was simulated by Monte Carlo Simulation with a mixed spin model of spin-$2$ and spin-$5/2$ on a layered honeycomb structure with nearest neighbour interactions to clarify the effects of interlayer interactions and single-ion anisotropies on compensation \cite{Nakamura}. In the last decade, the ferrimagnetic trilayered structure,\\ $Fe_{3}O_{4} (25nm)/Mn_{3}O_{4} (50nm)/Fe_{3}O_{4} (25nm)$ , prepared by \textit{oxygen-plasma-assisted} MBE, was shown to have magnetic compensation due to the formation of domain-wall-like configurations,
mainly in $Fe_{3}O_{4}$ \cite{Lin}. \\

%%%%%%%%%%%%%%%%%%%%%%%%%%%%%%%%%%%%%%%%%%%%
%Equilibrium Studies
%%%%%%%%%%%%%%%%%%%%%%%%%%%%%%%%%%%%%%%%%%%%
To examine compensation, numerical studies on the equilibrium (field-free and in the presence of static fields) properties of layered Ising ferrimagnets on various lattice geometries have been performed \cite{Oitmaa,Lv,Fadil,Diaz1,Diaz2,Chandra1,Chandra2,Chandra3}. The trilayered ferrimagnetic spin-1/2 Ising superlattices on square sublayers of ABA and AAB type [Figure \ref{fig_lattice_structure}] in the current study have an advantage as, in field-free cases they show compensation effect \cite{Diaz1,Diaz2,Chandra1,Chandra2,Chandra3}, even without site-dilution or mixed-spin structures. So, they are among the simplest systems to display compensation. The magnetic description provided by traditional Monte Carlo Simulation is in good agreement with the description provided by Inverse absolute of reduced residual magnetisation (IARRM) and Temperature interval between Critical and Compensation temperatures (TICCT) \cite{Chandra1,Chandra2,Chandra3} for both types of sandwiched configurations with square and triangular monolayers. The equilibrium studies are now well established.\\

However, real systems cannot preserve the pristine character, and disorder is almost unavoidable in the description of any spin model. The effect of spin-0 impurities (a kind of static disorder) on compensation in trilayered ferrimagnets (with triangular monolayers) is numerically investigated in a recent article \cite{Chandra4}. The disorder for systems studied in this work can as well be time-dependent. A few sources of time-dependent disorders \cite{Chandra5} are: (a) time varying interaction strength between pairs of spins; (b) the number of interacting spins for a particular site may be changing with time; (c) Even the nature of the spins (magnitude of spin, different magnetic atoms etc.) in the ordered structure may vary with time. As a result, modelling such temporally variable compositional and morphological disorders is extremely difficult. In an attempt in this direction \cite{Chandra5}, the trilayered spin$-1/2$ superlattices has been subjected to a uniform random external magnetic field with spatio-temporal variations. So the dynamic Hamiltonian for these systems then is a Random Field Ising Model (RFIM). RFIM was developed by Larkin in 1970 \cite{Larkin} and is historically used to model many remarkable static and dynamic behaviours in disordered systems \cite{Belanger}. Prominent examples of experimentally observed Random field type phenomenology in disordered systems range from disorder-induced frustration and electronic transport in disordered insulators to melting of intercalates in layered compounds e.g. $TiS_{2}$, \cite{Efros,Childress,Maher,Pastor,Kirkpatrick,Fisher1,Fisher2,Suter} to name a few. The simulational results and subsequent analyses in \cite{Sethna} and references therein explain how RFIM may describe various types of noises in magnets.\\

For a large variety of random field distributions, the critical exponents of the power laws are independent of the particular choice \cite{Sethna}. But it is yet to be established how a change in the nature of the continuous random external field, from uniform random \cite{Chandra5} to Gaussian random, affects the compensation phenomenon in the superlattices of Figure \ref{fig_lattice_structure}. Uniform random field has a lower and upper cut-off whereas the Gaussian random field admits all the possible real values of the field. Evidently, only one physical measure i.e. the standard deviation of these continuous random field distributions is common to both of these and provides us with a description of randomness, irrespective of the nature of the distribution. Thus the objective of the current study is to examine the influence of the Gaussian random external magnetic field (or, more specifically, the standard deviation of the Gaussian distribution) with certain spatiotemporal variation on the compensation phenomenon associated with a trilayered spin-$1/2$ Ising ferrimagnet with square monolayers.
%%%%%%%%%%%%%%%%%%%%%%%%%%%%%%%%%%%%%%%%%%%%
%Plan of the paper:
%%%%%%%%%%%%%%%%%%%%%%%%%%%%%%%%%%%%%%%%%%%%
The plan of the paper follows. The layered magnetic model and the dynamic Hamiltonian are described, in detail, in Section \ref{sec_model}. The simulational details are described in Section \ref{sec_simulation}. Section \ref{sec_results} contains the numerical results and associated discussions. In Section \ref{sec_summary}, the summary of the work is presented.

%%%%%%%%%%%%%%%%%%%%%%%%%%%%%%%%%%%%%%%%%%%%%%
\section{Outline of the Model}
\label{sec_model}
%%%%%%%%%%%%%%%%%%%%%%%%%%%%%%%%%%%%%%%%%%%%%%

\begin{figure*}[!htb]
	\begin{center}
		\begin{tabular}{c}
			\resizebox{7.5cm}{!}{\includegraphics[angle=0]{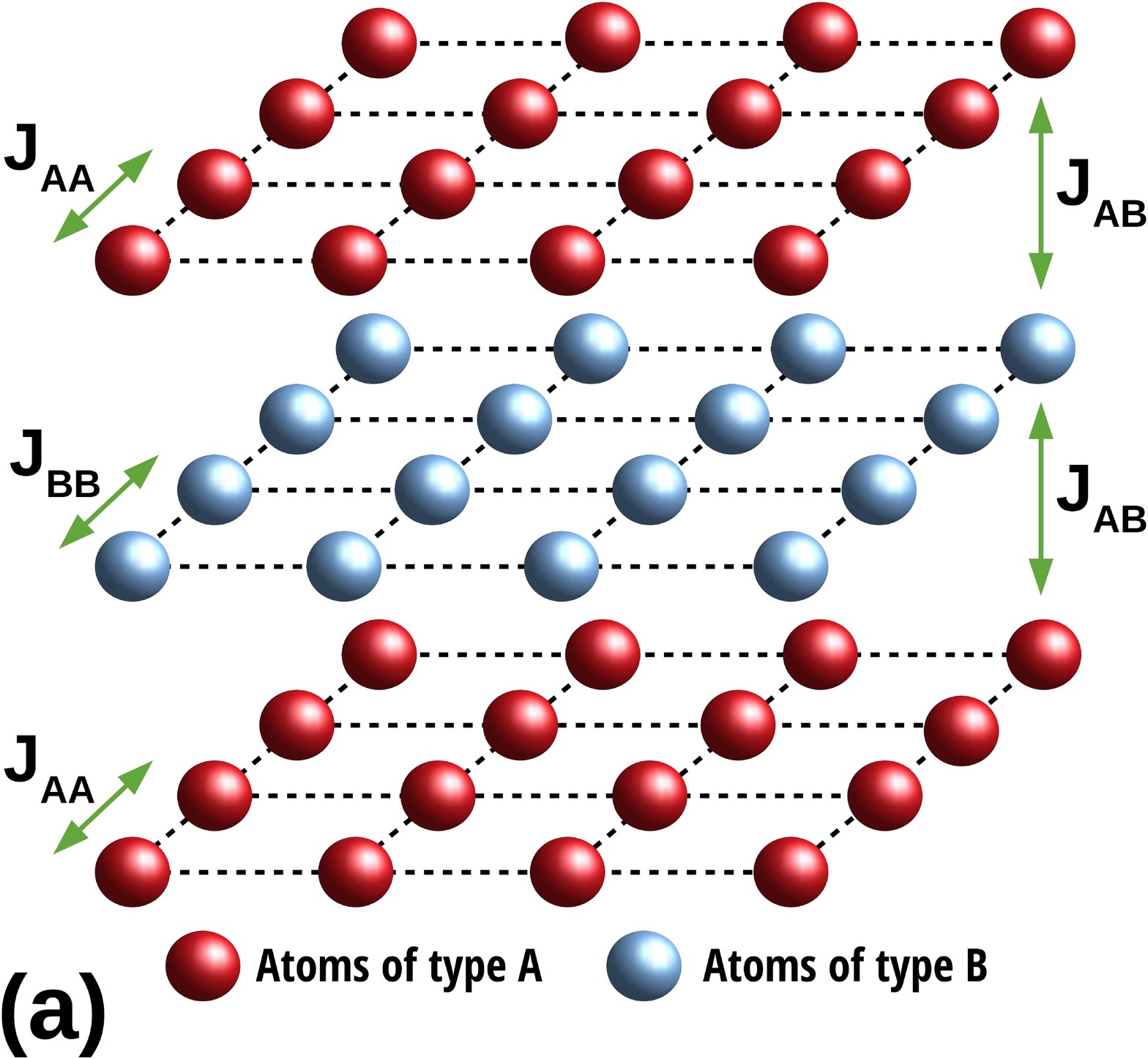}}
			{\hskip 1.5cm}
			\resizebox{7.5cm}{!}{\includegraphics[angle=0]{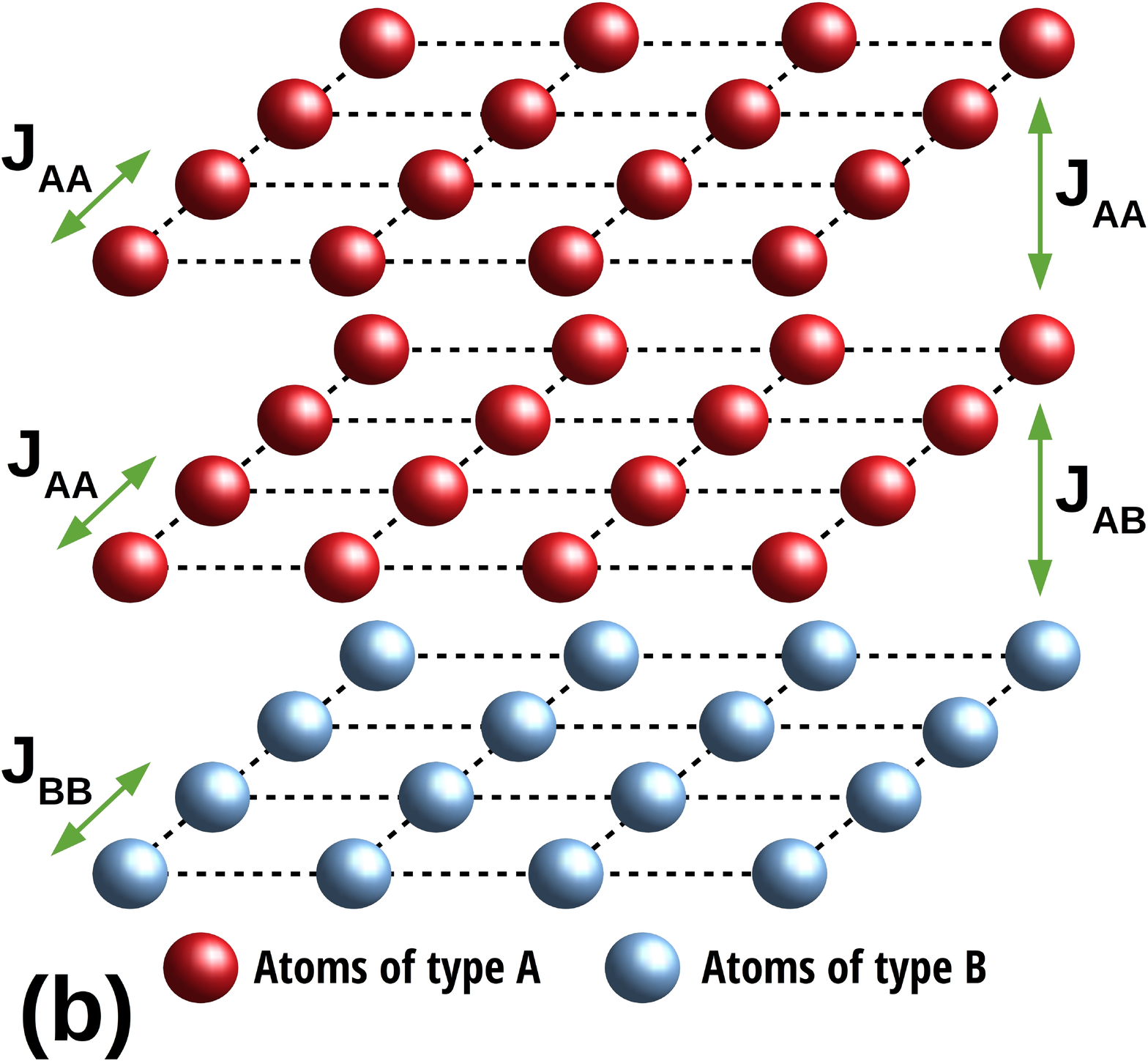}}
		\end{tabular}
		\caption{ (Colour Online) Miniaturised versions ($3\times4\times4$) of (a) ABA and (b) AAB square trilayered ferrimagnet with two types of theoretical atoms, $A$ and $B$. Each of the sublattices of the ferrimagnetic systems are formed on square lattice. The actual simulation is carried out on a system with $N_{sites}=3\times100\times100$ . Courtesy: \cite{Chandra5}}
		\label{fig_lattice_structure}
	\end{center}
\end{figure*}

The ferrimagnetic Ising superlattice in this study is similar to the one used in \cite{Chandra5}. Each site has spin value, $s=1/2$, and contains three magnetic sub-layers on square lattice. Each alternate layer is exhaustively composed of by either A or B type of atoms. The magnetic atoms on the top and bottom layers do not interact. [Fig.-\ref{fig_lattice_structure}]. The magnetic interaction between the like atoms (A-A and B-B) is ferromagnetic and between dislike atoms (A-B) is antiferromagnetic. Additionally to the cooperative interactions, the $z$-component of spins, $S_{i}^{z}$ at each site couples with a longitudinal Gaussian random external magnetic field, $h_{i}(t)$. At a particular site, this external field varies in time and at any time instant, the values of this local field are different over the lattice sites.
	
The spins interact Ising-like, limited to the nearest neighbours only, in-plane as well as inter-plane. Recent discoveries of $CrGeTe_{3}$ \cite{Gong}, $CrI_{3}$ \cite{Huang,Song} and $FeX_{2}(X=Cl,Br,I)$ \cite{McGuire} show the nearest-neighbour Ising interactions in few-layer limits of a magnetic material to be a reality. The time dependent Hamiltonian for such a trilayered ferrimagnetic system is:
\begin{eqnarray}
\nonumber
& &H(t) = - 
J_{11}\sum_{<t,{t}^{\prime }>}S_{t}^{z}S_{{t}^{\prime }}^{z} - J_{22}\sum_{<m,{m}^{\prime }>}S_{m}^{z}S_{{m}^{\prime }}^{z} \\\nonumber
& - &
J_{33} \sum_{<b,{b}^{\prime }>}S_{b}^{z}S_{{b}^{\prime }}^{z} - 
J_{12} \sum_{<t,m>}S_{t}^{z}S_{m}^{z} - 
J_{23} \sum_{<m,b>}S_{m}^{z}S_{b}^{z} \\
\label{eq_Hamiltonian}
&-& \sum_{i}h_{i}(t)S_{i}^{z}
\end{eqnarray}
$\langle t,{t}^{\prime }\rangle$, $\langle m,{m}^{\prime }\rangle$, $\langle b,{b}^{\prime }\rangle$ are nearest-neighbor pairs in the top, mid and bottom layers respectively and $\langle t,m\rangle$, $\langle m,b\rangle$ are, respectively, pairs of nearest-neighbor sites in, top \& mid and mid \& bottom layers. The first three terms are for the intra-planar ferromagnetic interactions. The fourth and fifth terms are for the inter-planar nearest neighbour interactions, between top and mid layers and mid and bottom layers, respectively. The sixth term denotes the spin-field interaction term of all the spins to the external Gaussian random magnetic field, at time instant $t$. Because of the interactions: $J_{AA}>0$ , $J_{BB}>0$, and $J_{AB}<0$. For an \textbf{ABA type system}: $J_{11}=J_{33}=J_{AA}$; $J_{22}=J_{BB}$ and $J_{12}=J_{23}=J_{AB}$. For an \textbf{AAB type system}: $J_{11}=J_{22}=J_{12}=J_{AA}$; $J_{33}=J_{BB}$ and $J_{23}=J_{AB}$.

\indent The local, Gaussian random external magnetic field values $h_{i}(t)$ at any site, $i$ at time instant $t$, is drawn from the following probability distribution:
\begin{equation}
\label{eq_prob_dist_Gaussian}
P_{Gaussian}(h_{i}(t)) = 
\frac{1}{\sqrt{2\pi \sigma^{2}}} \exp \left( \frac{-h_{i}^{2}(t)}{2\sigma^{2}}\right) 	
\end{equation}

Box-Muller algorithm \cite{Box} is used to get a Gaussian distribution of zero mean and standard deviation, $\sigma$ . The simulational details of implementation and associated important characteristics of such a time-varying field are discussed, in detail, in Appendix \ref{app_char_field}. 

%%%%%%%%%%%%%%%%%%%%%%%%%%%%%%%%%%%%%%%%%%%%
%%%%%%%%%%%%%%%%%%%%%%%%%%%%%%%%%%%%%%%%%%%%
\section{Simulation Protocol}
\label{sec_simulation}
%%%%%%%%%%%%%%%%%%%%%%%%%%%%%%%%%%%%%%%%%%%%

The Metropolis single spin-flip algorithm \cite{Landau, Binder} is employed for simulation of the system. The three square monolayers has $L^{2}$ sites with $L=100$. The $z$-components of spin projections of nearest neighbours, $S_{i}^{z}$ $(S_{i}^{z}=\pm1)$ contribute to the cooperative and spin-field interactions. At each site $i$, a local, time-varying, Gaussian random field, $h_{i}$ couples with each spin. In \cite{Diaz2}, Compensation temperature has been found to be practically constant for $L\geqslant60$, for the systems of this study in field-free study. From Appendix \ref{app_tcomp_size_field}, we see the compensation point is still independent of the system size in the vicinity of $L=100$, in presence of the external Gaussian field of this study. So the chosen system size is statistically reliable for simulational investigation. For both the configurations, the systems are initiated at a high temperature paramagnetic phase, with randomly selected half of the total spin projections being ``UP" (with $S_{i}^{z}=+1$) and the rest being ``DOWN" (with $S_{i}^{z}=-1$) (Using $1$ for $1/2$ fixes up the energy scale). At a fixed temperature $T$, the spin flipping is governed by the Metropolis rate \cite{Metropolis, Newman}, of Equation [\ref{eq_metropolis}]:
\begin{equation}
\label{eq_metropolis}
P(S_{i}^{z} \to -S_{i}^{z}) = \text{min} \{1, \exp (-\Delta E/k_{B}T)\}
\end{equation}
where the associated change in internal energy in flipping the $i$-th spin projection from $S_{i}^{z}$ to $-S_{i}^{z}$, is $\Delta E$ . Similar $3L^{2}$ individual, random single-spin updates constitute one Monte Carlo sweep (MCS) of the entire system (unit of time in this study). Periodic boundary conditions in-plane and Open boundary conditions along the vertical are employed.

The systems are kept for $10^{5}$ MCS at every temperature step. The last \textbf{equilibrium} configuration at the previous \textit{higher} temperature acts as the starting configuration at a new \textit{lower} temperature. For the first cumulative $5\times10^{4}$ MCS: the system is allowed to equilibrate in the field-free environment first and then the system attains steady-state in the presence of the external field (the provided time, for attaining eqilibrium and steady state, is sufficient [Refer to (a) Figures \ref{fig_ABA_field_mag_time} \& \ref{fig_AAB_field_mag_time} and discussions therein and (b) Appendix \ref{app_transient}]). After that the external field is kept switched on for the next $5\times10^{4}$ MCS. So for the systems, the exposure time interval in the field, $\delta$ is $5\times10^{4}$. The temperatures are measured in units of $J_{BB}/k_{B}$. For each of the fixed standard deviation of the Gaussian random field, the system is observed for seven equidistant values of $J_{AA}/J_{BB}$, from $0.04$ to $1.0$ with an interval of $0.16$. For each fixed value of $J_{AA}/J_{BB}$, $J_{AB}/J_{BB}$ is decreased from $-0.04$ to $-1.0$ with a step of $-0.16$.\\
\indent For any combination of $J_{AA}/J_{BB}$ and $J_{AB}/J_{BB}$, and a fixed standard deviation of the external field $\sigma$, the time averages of the following quantities are calculated after equilibration at any temperature, $(T)$ in the following manner \cite{Robb,Chandra5}:\\
\textbf{(1) Sublattice magnetisations} are calculated at time instant say, $t$, by:
\begin{equation}
M_{q}(T,t)=\frac{1}{L^{2}}\sum_{x,y=1}^{L} \left( S_{q}^{z}(T,t)\right)_{xy}
\end{equation}
The time averaged sublattice magnetizations is calculated by:
\begin{equation}
\langle M_{q}(T)\rangle =\frac{1}{\delta}\int_{t_{0}}^{t_{0}+\delta} M_{q}(T,t)dt
\end{equation}
where $q$ is to be replaced by $t,m\text{ or }b$ for top, mid and bottom layers.\\
\textbf{(2) The order parameter}, $O(T)$, for the trilayer at temperature, $T$ is defined as:
\begin{equation}
O(T)=\frac{1}{3}(\langle M_{t}(T)\rangle+\langle M_{m}(T)\rangle+\langle M_{b}(T)\rangle)
\end{equation}
\\
\textbf{(3) Fluctuation of the order parameter,} $\Delta O(T)$ at temperature, $T$ as follows:
\begin{equation}
{\Delta O}(T)=\sqrt{\dfrac{1}{\delta} \int_{t_{0}}^{t_{0}+\delta} \left[M(T,t)-O(T)\right]^{2}dt }
\end{equation}
where $M(T,t)$ is the total magnetisation of the whole system, at temperature, $T$, calculated at the $t$-th time instant. \\
\textbf{(4) The time averaged value of cooperative energy per site},  $\langle E(T) \rangle$, at temperature, $T$, is determined by:
\begin{eqnarray}
\nonumber 
& \langle E (T)\rangle_{ABA} & = \dfrac{-1}{3L^{2}\delta}  \int_{t_{0}}^{t_{0}+\delta} dt [ J_{AA}(\sum_{<t,{t}^{\prime }>}S_{t}^{z}S_{{t}^{\prime }}^{z} 
\\\nonumber 
&+& \sum_{<b,{b}^{\prime }>}S_{b}^{z}S_{{b}^{\prime }}^{z}) + J_{BB} \sum_{<m,{m}^{\prime }>}S_{m}^{z}S_{{m}^{\prime }}^{z} 
\\
&+& J_{AB} (\sum_{<t,m>}S_{t}^{z}S_{m}^{z}+\sum_{<m,b>}S_{m}^{z}S_{b}^{z})] 
\end{eqnarray}
and
\begin{eqnarray}
\nonumber 
& \langle E (T)\rangle_{AAB} & = \dfrac{-1}{3L^{2}\delta}  \int_{t_{0}}^{t_{0}+\delta} dt [ J_{AA}(\sum_{<t,{t}^{\prime }>}S_{t}^{z}S_{{t}^{\prime }}^{z}
\\\nonumber 
&+& \sum_{<m,{m}^{\prime }>}S_{m}^{z}S_{{m}^{\prime }}^{z} \sum_{<t,m>}S_{t}^{z}S_{m}^{z} ) \\
&+& J_{BB} \sum_{<b,{b}^{\prime }>}S_{b}^{z}S_{{b}^{\prime }}^{z} + J_{AB} \sum_{<m,b>}S_{m}^{z}S_{b}^{z} ]
\end{eqnarray}
\textbf{(5) The fluctuation of the cooperative energy per site} at temperature, $T$, by:
\begin{equation}
{\Delta E}(T)=\sqrt{\dfrac{1}{\delta} \int_{t_{0}}^{t_{0}+\delta} \left[E(T,t)-\langle E (T)\rangle\right]^{2}dt }
\end{equation}
where $E(T,t)$ is the instantaneous cooperative energy per site, for the system at time $t$ and temperature, $T$, within the exposure interval, $\delta$. 

At the pseudo-critical temperatures, the fluctuations peak. Around this temperature close range simulations were performed to narrow down the position of the reported critical temperatures with an accuracy of, $\Delta T_{crit}=0.04$ . Compensation temperature ($<T_{crit}$), where the total magnetisation again becomes zero, is determined by linear interpolation from the two neighbouring points across the zero of magnetization in the plots of order parameter versus temperature [e.g. Figure \ref{fig_mag_fr_afr}(a)]. The upper bounds of linear interpolation provide us with an estimate of the errors with the values of compensation points \cite{Scarborough}. The Jackknife method \cite{Newman} is used to provide an estimate of the errors with the magnetizations and fluctuations. 
%%%%%%%%%%%%%%%%%%%%%%%%%%%%%%%%%%%%%%%%%%%%%

\section{Results and discussions}
\label{sec_results}
%%%%%%%%%%%%%%%%%%%%%%%%%%%%%%%%%%%%%%%%%%%%%
\subsection{Thermodynamic Response}
\label{subsec_response}
%%%%%%%%%%%%%%%%%%%%%%%%%%%%%%%%%%%%%%%%%%%%%
\subsubsection{Magnetization versus temperature:}
\label{subsubsec_magvtemp}
%%%%%%%%%%%%%%%%%%%%%%%%%%%%%%%%%%%%%%%%%%%%%

\begin{figure*}[!htb]
	\begin{center}
		\begin{tabular}{c}
			
			\resizebox{8.5cm}{!}{\includegraphics[angle=0]{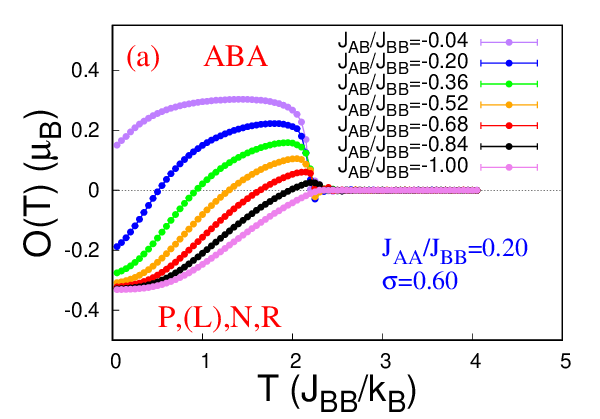}}
			\resizebox{8.5cm}{!}{\includegraphics[angle=0]{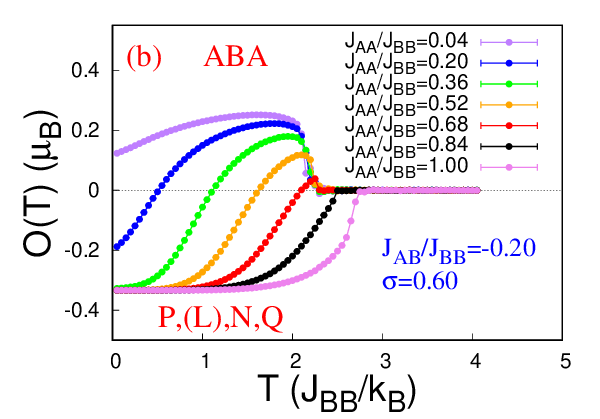}}\\
			
			\resizebox{8.5cm}{!}{\includegraphics[angle=0]{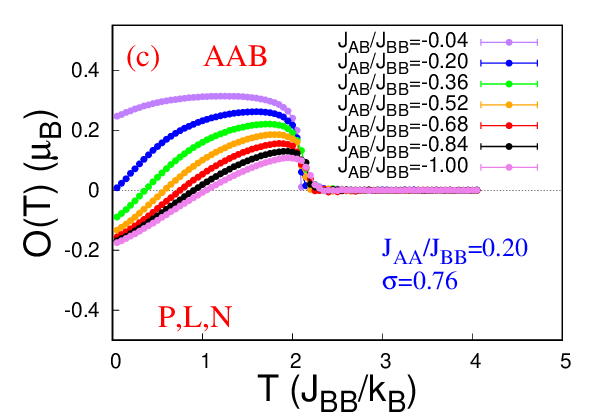}}
			\resizebox{8.5cm}{!}{\includegraphics[angle=0]{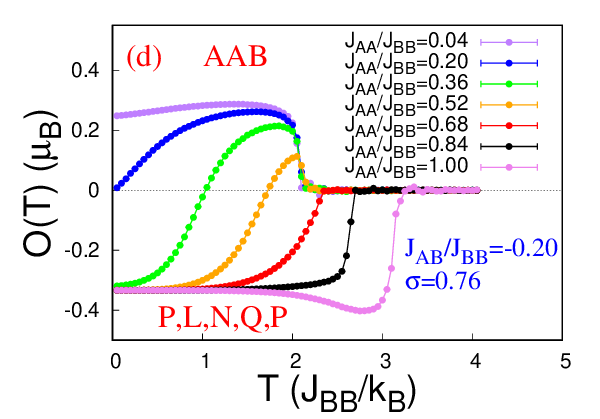}}
			
		\end{tabular}
		\caption{ (Colour Online) Plots of Order parameter versus reduced temperature for a $3\times 100 \times 100$ system of type: (a) ABA with $J_{AA}/J_{BB}=0.20$ and variable $J_{AB}/J_{BB}$ for $\sigma=0.60$; (b) ABA with $J_{AB}/J_{BB}=-0.20$ and variable $J_{AA}/J_{BB}$ for $\sigma=0.60$; (c) AAB with $J_{AA}/J_{BB}=0.20$ and variable $J_{AB}/J_{BB}$ for $\sigma=0.76$; (d) AAB with $J_{AB}/J_{BB}=-0.20$ and variable $J_{AA}/J_{BB}$ for $\sigma=0.76$ . Shift of Compensation and Critical temperatures towards higher temperature end are witnessed with increase in any of the coupling ratios. The field-driven vanishing of compensation is witnessed for the weakest combination of coupling strengths.}
		\label{fig_mag_fr_afr}
	\end{center}
\end{figure*}

In the few cases in Figure \ref{fig_mag_fr_afr}, for a fixed standard deviation of the external field with characteristics of Section \ref{app_char_field}, we see the compensation and critical temperatures shift as we increase the magnitude of any of the coupling strengths. As we increase the magnitude of either of the coupling strengths, we can identify the nature of the magnetization curves by the N\`{e}el classification scheme. A detailed discussion on the classification schemes (e.g. P-type, N-type, R-type etc.) can be found out \cite{Neel,Chikazumi,Strecka}. \textbf{For the ABA configuration}: \textbf{in Figure \ref{fig_mag_fr_afr} (a)}  with $J_{AA}/J_{BB}=0.20$ and $\sigma=0.60$: for $J_{AB}/J_{BB}=-0.04$ we see a P-type magnetization; all the intermediate curves are of N-type and for $J_{AB}/J_{BB}=-1.00$ we see a R-type magnetization; and \textbf{in Figure \ref{fig_mag_fr_afr} (b)} with $J_{AB}/J_{BB}=-0.20$ and $\sigma=0.60$: for $J_{AA}/J_{BB}=0.04$ we see a P-type magnetization; the intermediate curves up to are of N-type and for $J_{AB}/J_{BB}=1.00$ we see a Q-type magnetization. The L-type, within braces and not explicitly shown, would be encountered while moving from the former type to the latter. For the weakest combination of coupling strengths, we witness the field-driven vanishing of compensation. \textbf{For the AAB configuration}: \textbf{in Figure \ref{fig_mag_fr_afr} (c)}  with $J_{AA}/J_{BB}=0.20$ and $\sigma=0.76$: for $J_{AB}/J_{BB}=-0.04$ we see a P-type magnetization and for $J_{AB}/J_{BB}=-0.20$ we see an L-type magnetization and all other curves are of N-type and \textbf{in Figure \ref{fig_mag_fr_afr} (d)} with $J_{AB}/J_{BB}=-0.20$ and $\sigma=0.76$: for $J_{AA}/J_{BB}=0.04$ we see a P-type magnetization; for $J_{AA}/J_{BB}=0.20$ we see an L-type; the $J_{AA}/J_{BB}=0.36,0.52$ curves are of N-type; the $J_{AA}/J_{BB}=0.68,0.84$ curves are of Q-type and for $J_{AB}/J_{BB}=1.00$ we see a P-type magnetization again. For the two weakest combinations of coupling strengths, we again witness the field-driven vanishing of compensation in the AAB configuration. The magnetic response under the influence of Gaussian random magnetic field with spatio-temporal variation is quite similar to what we see for the spatio-temporally varying uniform random magnetic field in \cite{Chandra5}.\\

\begin{figure*}[!htb]
	\begin{center}
		\begin{tabular}{c}
			
			\resizebox{8.5cm}{!}{\includegraphics[angle=0]{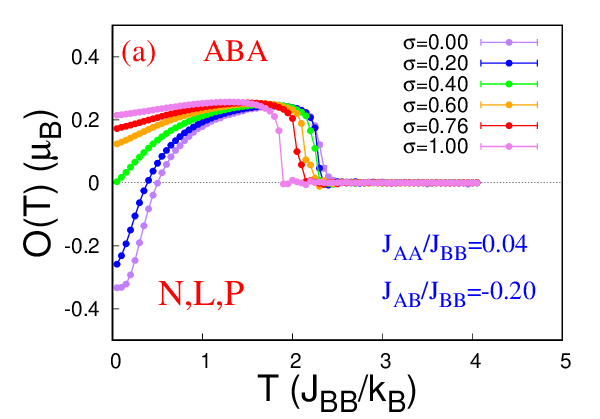}}
			\resizebox{8.5cm}{!}{\includegraphics[angle=0]{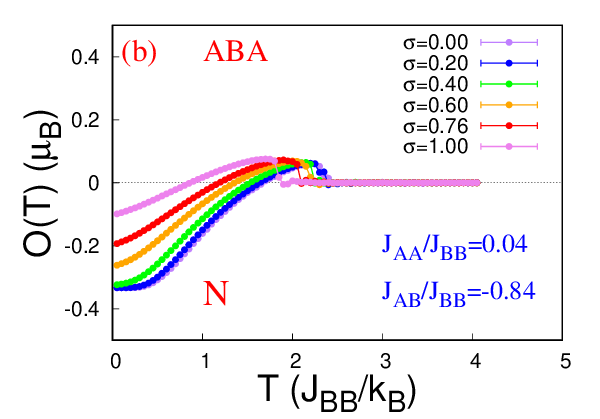}}\\
			
			\resizebox{8.5cm}{!}{\includegraphics[angle=0]{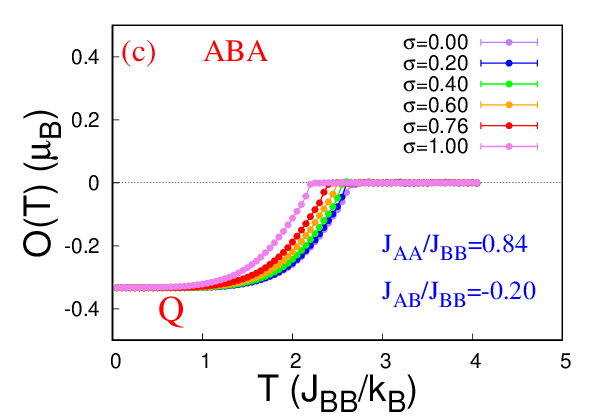}}
			\resizebox{8.5cm}{!}{\includegraphics[angle=0]{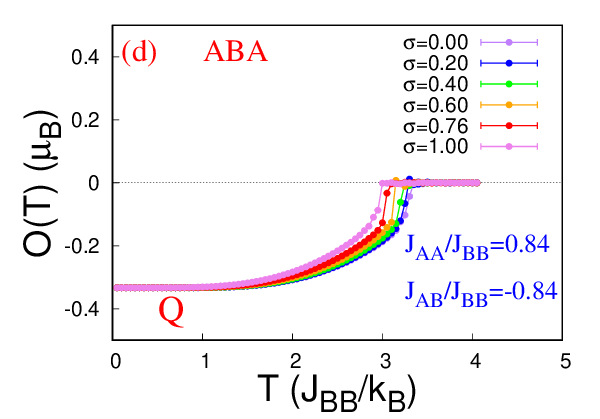}}\\
			
		\end{tabular}
		\caption{ (Colour Online) Magnetic response of the ABA trilayered ($3\times100\times100$) system. The shift of both, the compensation (where it is present) and critical temperatures towards the low temperature ends and shift of the magnetic behaviours between N,L,P,Q etc. type of ferrimagnetism, with increase in the standard deviation of the uniform random external magnetic field, are clearly visible in all these plots. The type L within brackets is explicitly not seen in the plots but encountered in-transition.  Where, the errorbars are not visible, they are smaller than the area of the point-markers. }
		\label{fig_mag_response_ABA}
	\end{center}
\end{figure*}

\begin{figure*}[!htb]
	\begin{center}
		\begin{tabular}{c}
			
			\resizebox{8.5cm}{!}{\includegraphics[angle=0]{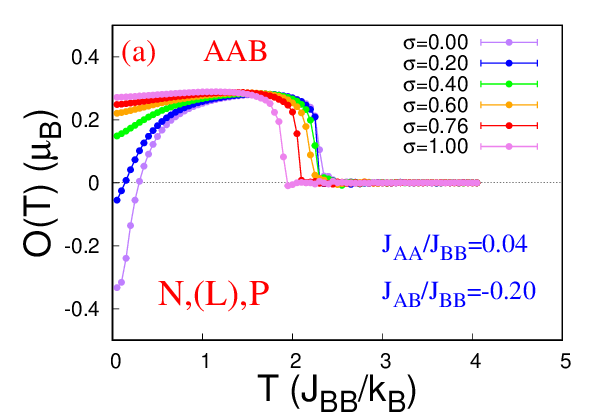}}
			\resizebox{8.5cm}{!}{\includegraphics[angle=0]{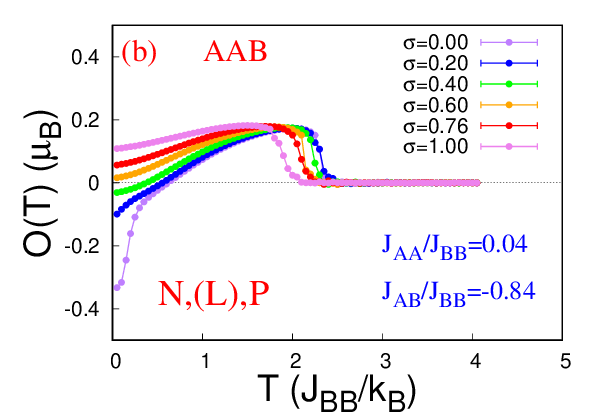}}\\
			
			\resizebox{8.5cm}{!}{\includegraphics[angle=0]{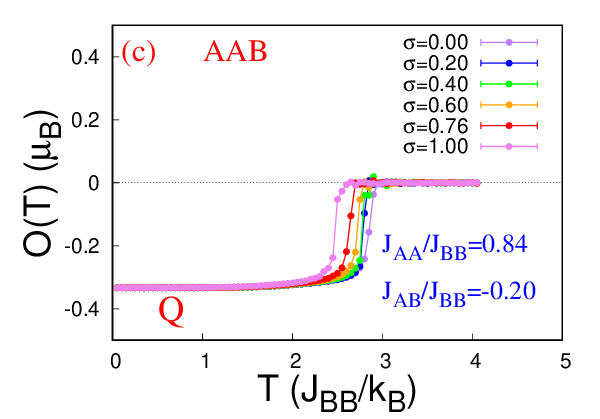}}
			\resizebox{8.5cm}{!}{\includegraphics[angle=0]{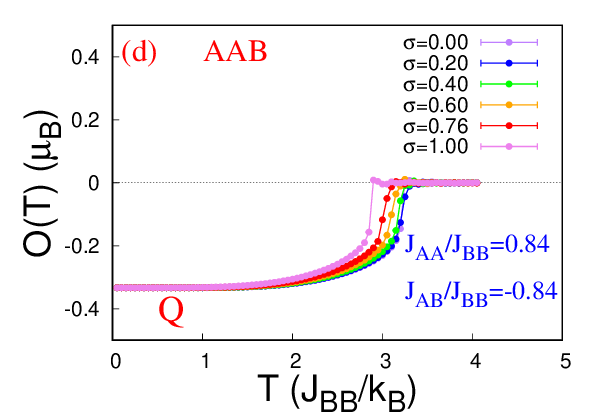}}
		\end{tabular}
		\caption{ (Colour Online) Magnetic response of the AAB trilayered ($3\times100\times100$) system. The shift of both, the compensation (where it is present) and critical temperatures towards the low temperature ends and shift of the magnetic behaviours between N,L,P,Q etc. type of ferrimagnetism, with increase in the standard deviation of the uniform random external magnetic field, are clearly visible in all these plots. The type L within brackets is explicitly not seen in the plots but encountered in-transition. Where, the errorbars are not visible, they are smaller than the area of the point-markers.}
		\label{fig_mag_response_AAB}
	\end{center}
\end{figure*}

Now we will focus on the effects of the randomness of the external Gaussian random field has on the magnetic response. For any fixed combination of the coupling strengths, an increase in the value of the standard deviation of the external Gaussian random field decreases the compensation and critical temperatures for both the ABA and AAB type of sandwiched structures [Figures \ref{fig_mag_response_ABA} and \ref{fig_mag_response_AAB}]. Similar to the uniform random field \cite{Chandra5}, as we increase the randomness of the external Gaussian random field the decrement for the compensation temperatures is much more visible than the decrement of critical temperature, with or without compensation. The \textit{field driven absence of compensation phenomenon} is also present in Figure \ref{fig_mag_response_ABA}(a) for ABA and Figures \ref{fig_mag_response_AAB}(a)\&(b) for AAB configration. Like the uniform random external field, we can see and identify the nature of ferrimagnetic curves and field-driven transitions among them in Figures \ref{fig_mag_response_ABA} and \ref{fig_mag_response_AAB}. According to the classification schemes of references \cite{Neel,Chikazumi,Strecka}: \textbf{(A) For ABA}, in Figure \ref{fig_mag_response_ABA}(a), the magnetic response changes from type-$N$ $(\sigma ={0,0.20})$ to type-$L$ $(\sigma =0.40)$ to type-$P$ $(\sigma ={0.60,0.76,1.00})$; In Figure \ref{fig_mag_response_ABA}(b), for all the fields only type-$N$ response is witnessed; and in Figures \ref{fig_mag_response_ABA}(c)\&(d), we see only type-$Q$ response for all the fields. \textbf{(B) For AAB}, in Figure \ref{fig_mag_response_AAB}(a), the transition happens from type-$N$ $(\sigma={0.00,0.20})$ to type-$P$ $(\sigma={0.40,0.60,0.76,1.00})$ via type-$L$; Similar transitions are witnessed in Figure \ref{fig_mag_response_AAB}(b); in Figure \ref{fig_mag_response_AAB}(c) \& (d), all the magnetic responses are of type-$Q$.

%%%%%%%%%%%%%%%%%%%%%%%%%%%%%%%%%%%%%%%%%%%%%
\subsubsection{Fluctuations versus temperature:}
\label{subsubsec_flucvtemp}
\begin{figure*}[!htb]
	\begin{center}
		\begin{tabular}{c}
			
			\resizebox{8.5cm}{!}{\includegraphics[angle=0]{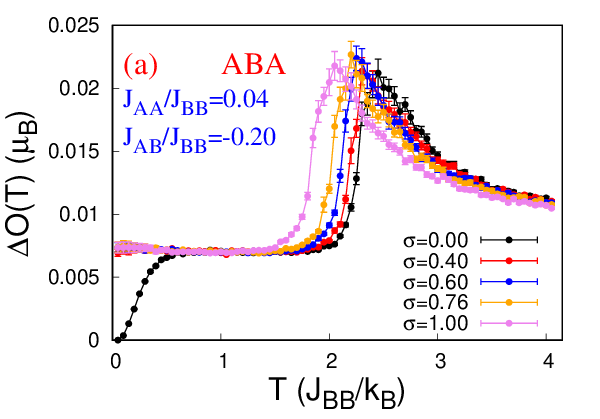}}
			
			\resizebox{8.5cm}{!}{\includegraphics[angle=0]{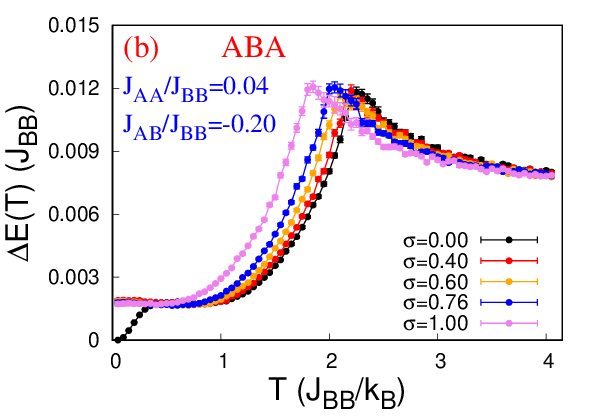}}\\
			
			\resizebox{8.5cm}{!}{\includegraphics[angle=0]{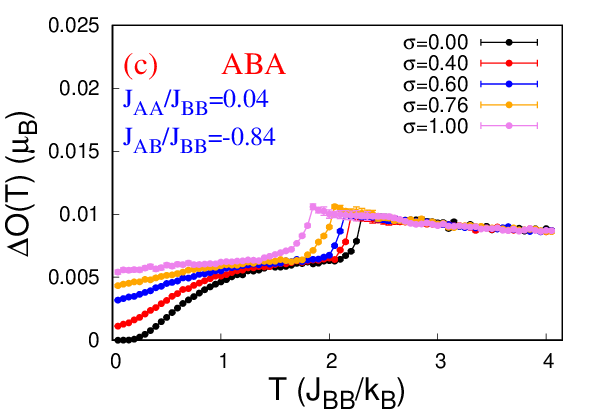}}
			
			\resizebox{8.5cm}{!}{\includegraphics[angle=0]{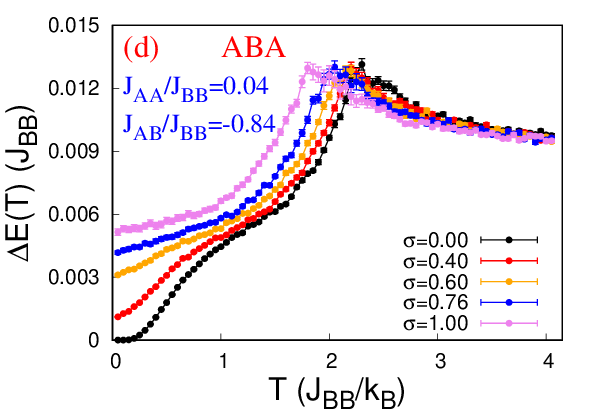}}\\
			
		\end{tabular}
		\caption{ (Colour Online) Temperature dependence of Fluctuation of order parameter, $\Delta O (T)$ and Fluctuation of cooperative energy per site, $\Delta E (T)$, for ABA type $(3\times100\times100)$ configuration in (a)-(d) with ${J_{AA}}/{J_{BB}}=0.04$ and ${J_{AB}}/{J_{BB}}=-0.20$ and with ${J_{AA}}/{J_{BB}}=0.04$ and ${J_{AB}}/{J_{BB}}=-0.84$. Where, the errorbars are not visible, they are smaller than the dimension of the point-markers. The nature of the curves prominently shows the shift of critical temperatures and even reason for absence of compensation can be understood from the low temperature segment of the curves.}
		\label{fig_fluc_mageng_ABA}
	\end{center}
\end{figure*}

\begin{figure*}[!htb]
	\begin{center}
		\begin{tabular}{c}

			\resizebox{8.5cm}{!}{\includegraphics[angle=0]{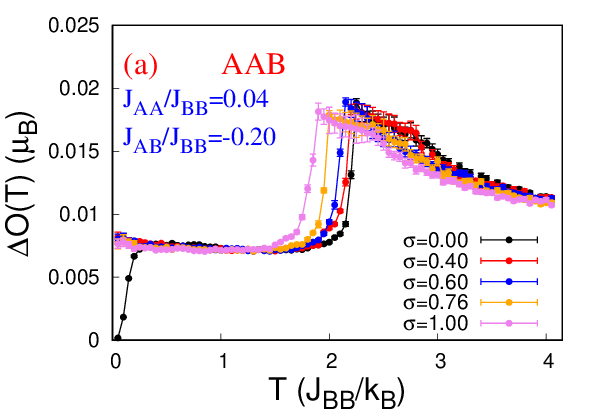}}
			
			\resizebox{8.5cm}{!}{\includegraphics[angle=0]{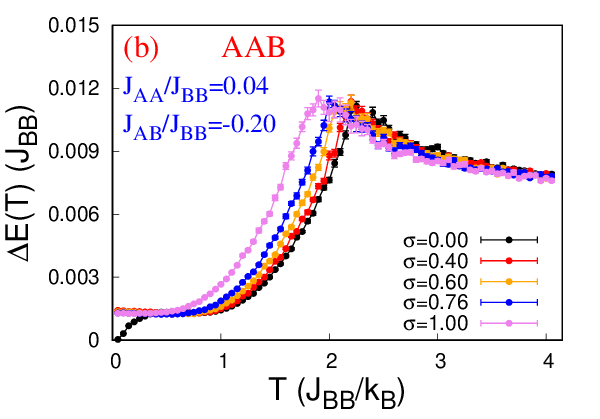}}\\
			
			\resizebox{8.5cm}{!}{\includegraphics[angle=0]{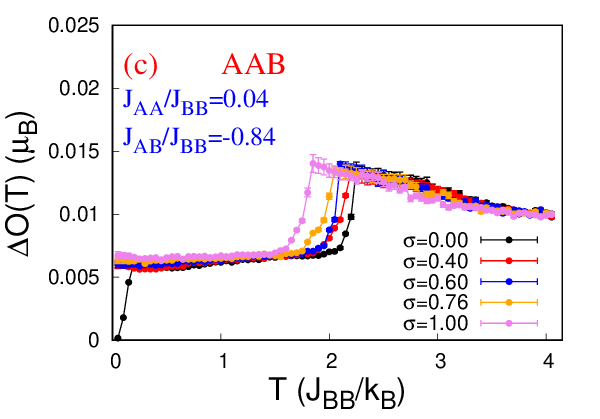}}
			
			\resizebox{8.5cm}{!}{\includegraphics[angle=0]{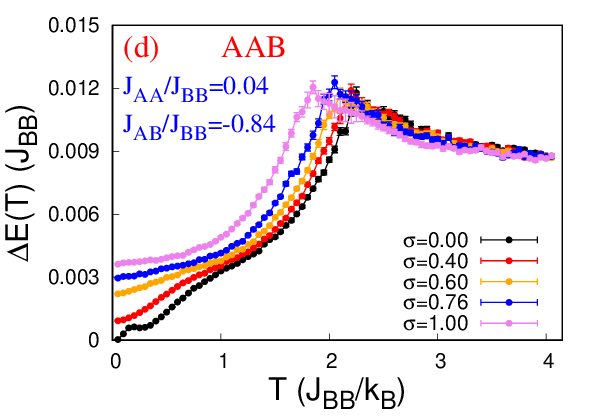}}
			
		\end{tabular}
		\caption{ (Colour Online) Temperature dependence of Fluctuation of order parameter, $\Delta O (T)$ and Fluctuation of cooperative energy per site, $\Delta E (T)$, for AAB type $(3\times100\times100)$ configuration in (a)-(d) with ${J_{AA}}/{J_{BB}}=0.04$ and ${J_{AB}}/{J_{BB}}=-0.20$ and with ${J_{AA}}/{J_{BB}}=0.04$ and ${J_{AB}}/{J_{BB}}=-0.84$. Where, the errorbars are not visible, they are smaller than the dimension of the point-markers. The nature of the curves prominently shows the shift of critical temperatures and even reason for absence of compensation can be understood from the low temperature segment of the curves.}
		\label{fig_fluc_mageng_AAB}
	\end{center}
\end{figure*}

%%%%%%%%%%%%%%%%%%%%%%%%%%%%%%%%%%%%%%%%%%%%%
To better understand (a) the shifts of compensation and critical temperatures and (b) the reason behind the field driven vanishing of compensation, we will now observe both the fluctuations: fluctuation of the order parameter and fluctuation of the cooperative energy per site, as functions of temperature while standard deviation of the external Gaussian random field acts as the parameter. We witness a plateau with a smeared peak in the vicinity of compensation point for both the fluctuations of order parameter and energy in Figures \ref{fig_fluc_mageng_ABA} and \ref{fig_fluc_mageng_AAB} for the ABA and AAB configurations respectively. We clearly see the compensation and critical temperatures moving towards lower temperature values, with the increase in the standard deviation of the Gaussian field. Even the smeared peaks at the low temperature segments flatten out as the standard deviation is increased in steps, which signifies the vanishing of compensation. So we again witness a field-driven vanishing of compensation. At the lower parts of the temperature axis, the increase in both the fluctuations imply that magnetic ordering is gradually decreasing with the increase of the standard deviation of the external field.   

\begin{figure*}[!htb]
	\begin{center}
		\begin{tabular}{c}

			$\mathbf{\text{ }ABA:\text{ } J_{AA}/J_{BB}=0.04;\text{ } J_{AB}/J_{BB}=-0.20 \text{ and } T=T_{comp}(\sigma=0)=0.502}$\\
			
			\phantom{}
			\hspace{0.00cm} \textbf{Top layer} \hspace{4.250cm} \textbf{Mid layer} \hspace{4.250cm} \textbf{Bottom layer}
			\\ 
			
			$\mathbf{(a)\text{ } \sigma=0.00}$\\
			
			\resizebox{6.0cm}{!}{\includegraphics[angle=0]{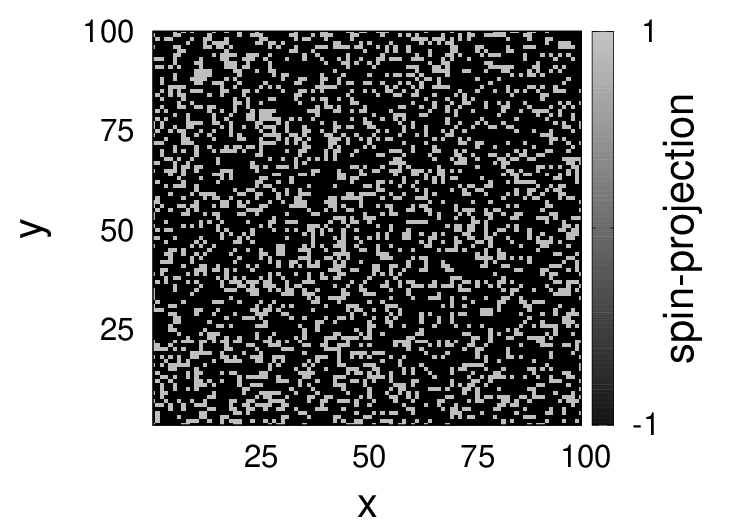}}
			
			\resizebox{6.0cm}{!}{\includegraphics[angle=0]{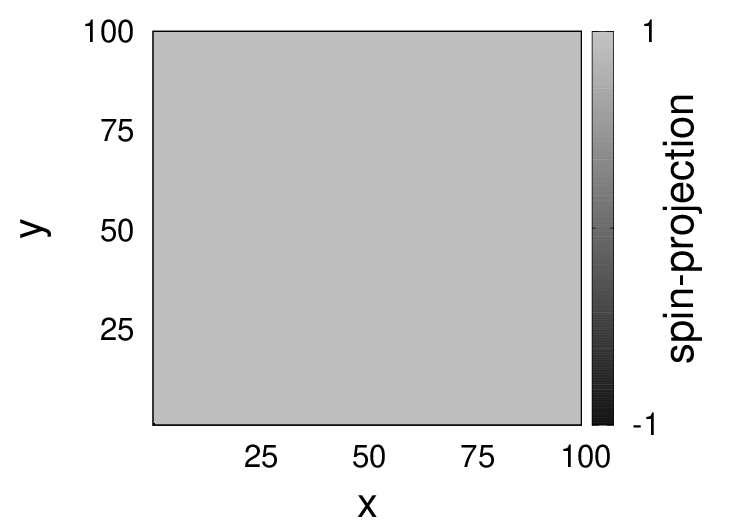}}
			
			\resizebox{6.0cm}{!}{\includegraphics[angle=0]{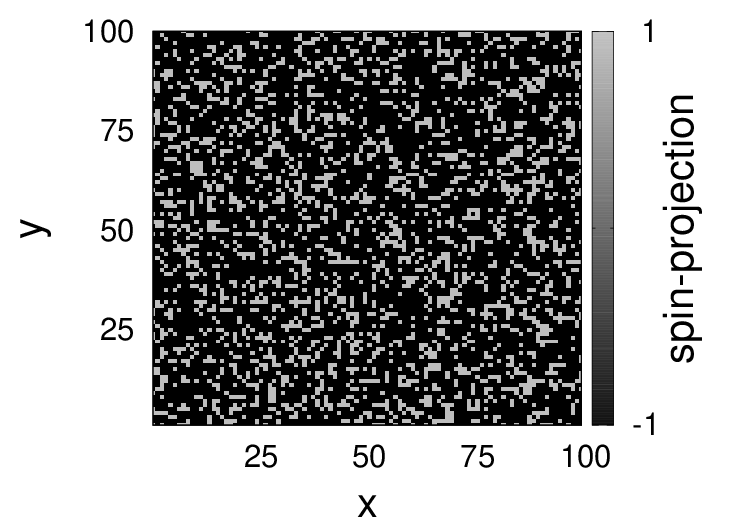}}\\
			
			\phantom{}
			\hspace{-0.00cm} $M_{t}(T,\sigma)=-0.484$ \hspace{3.00cm} $M_{m}(T,\sigma)=+1.000$ \hspace{2.75cm} $M_{b}(T,\sigma)=-0.502$\\
			
			\\

			$\mathbf{(b)\text{ } \sigma=0.20}$\\
			
			\resizebox{6.0cm}{!}{\includegraphics[angle=0]{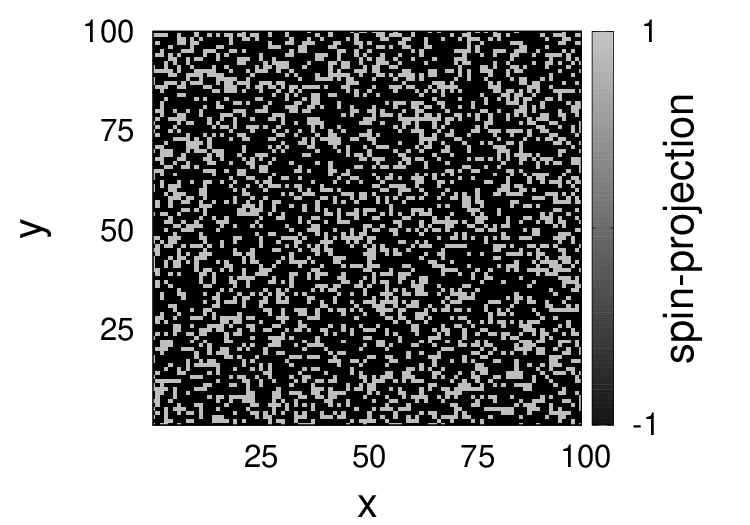}}
			
			\resizebox{6.0cm}{!}{\includegraphics[angle=0]{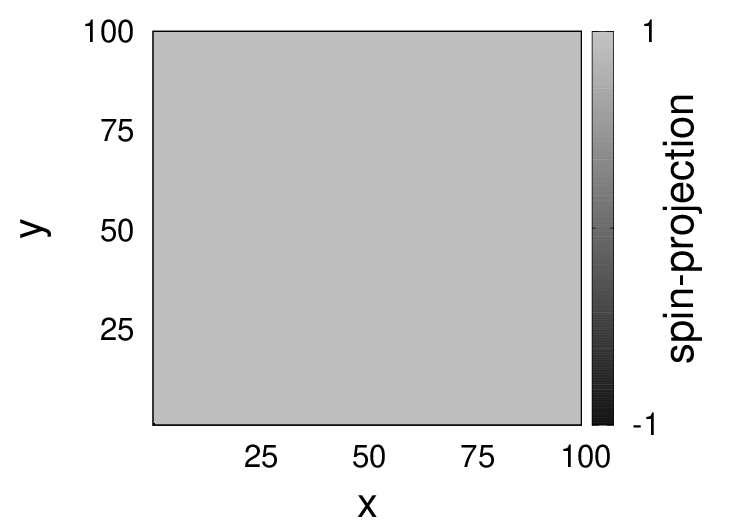}}
			
			\resizebox{6.0cm}{!}{\includegraphics[angle=0]{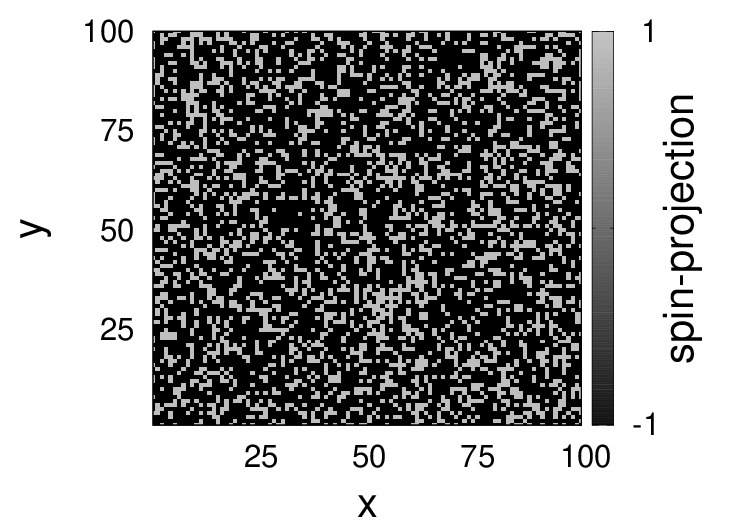}}\\
			
			\phantom{}
			\hspace{-0.00cm} $M_{t}(T,\sigma)=-0.394$ \hspace{3.00cm} $M_{m}(T,\sigma)=+1.000$ \hspace{2.75cm} $M_{b}(T,\sigma)=-0.398$\\
			
			\\
			
			$\mathbf{(c)\text{ } \sigma=0.76}$\\
			
			\resizebox{6.0cm}{!}{\includegraphics[angle=0]{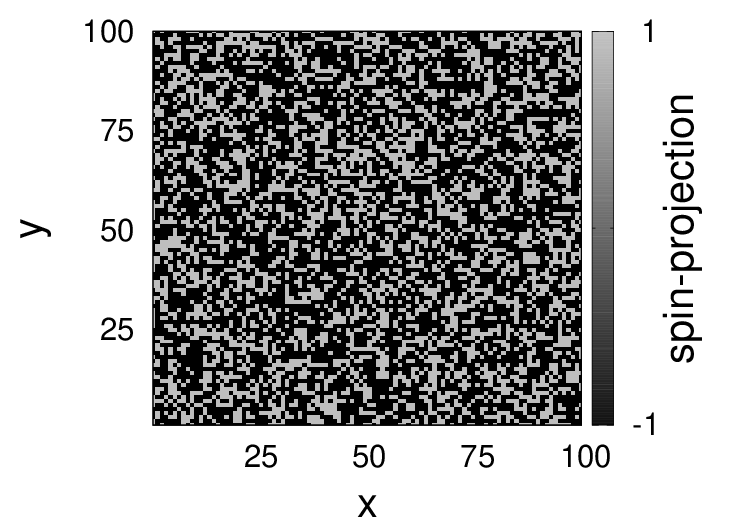}}
			
			\resizebox{6.0cm}{!}{\includegraphics[angle=0]{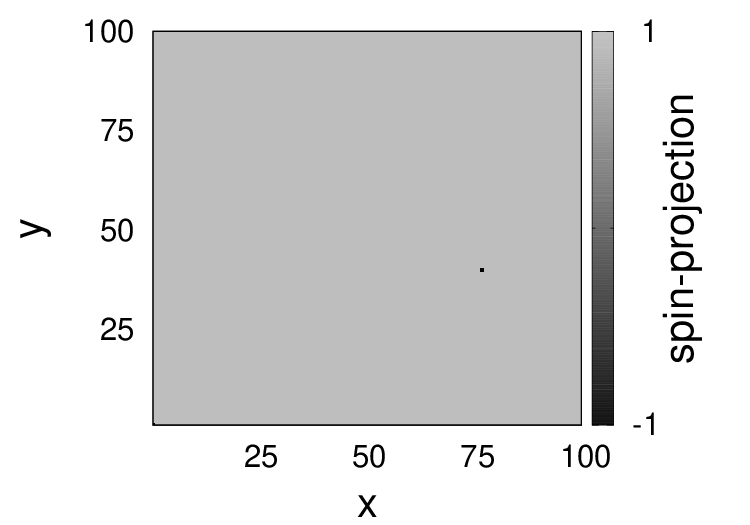}}
			
			\resizebox{6.0cm}{!}{\includegraphics[angle=0]{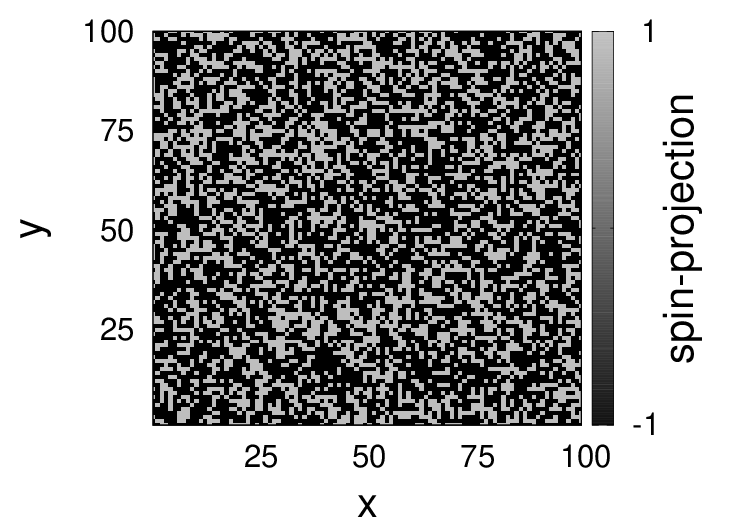}}\\
			
			\phantom{}
			\hspace{-0.00cm} $M_{t}(T,\sigma)=-0.201$ \hspace{3.00cm} $M_{m}(T,\sigma)=+1.000$ \hspace{2.75cm} $M_{b}(T,\sigma)=-0.192$\\
			
			\\
			
		\end{tabular}
		\caption{ \textbf{For ABA configuration}: Lattice morphologies of \textbf{top layer (at Left)}; \textbf{mid layer (at Middle)} and \textbf{bottom layer (at Right)} at $t=t_{morph}=10^{5}$ $MCS$ for $J_{AA}/J_{BB}=0.04$ and $J_{AB}/J_{BB}=-0.20$ and a few standard deviations of the external field. The magnetisations are rounded-off to three decimal places. The shift and vanishing of compensation in the respective cases (b) and (c) is due to the significant reduction of magnetic ordering in the top and bottom layers i.e. \textit{surface A-layers}.}
		\label{fig_morphology_ABA}
	\end{center}
\end{figure*}

\begin{figure*}[!htb]
	\begin{center}
		\begin{tabular}{c}

			$\mathbf{\text{ }AAB:\text{ } J_{AA}/J_{BB}=0.04;\text{ } J_{AB}/J_{BB}=-0.20 \text{ and } T=T_{comp}(\sigma=0)=0.296}$\\
			
			\phantom{}
			\hspace{0.00cm} \textbf{Top layer} \hspace{4.250cm} \textbf{Mid layer} \hspace{4.250cm} \textbf{Bottom layer} \\

			$\mathbf{(a)\text{ } \sigma=0.00}$\\
			
			\resizebox{6.0cm}{!}{\includegraphics[angle=0]{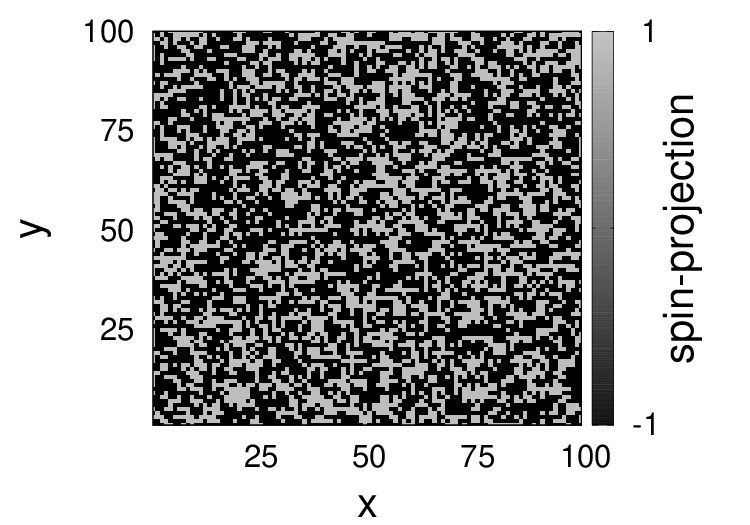}}
			
			\resizebox{6.0cm}{!}{\includegraphics[angle=0]{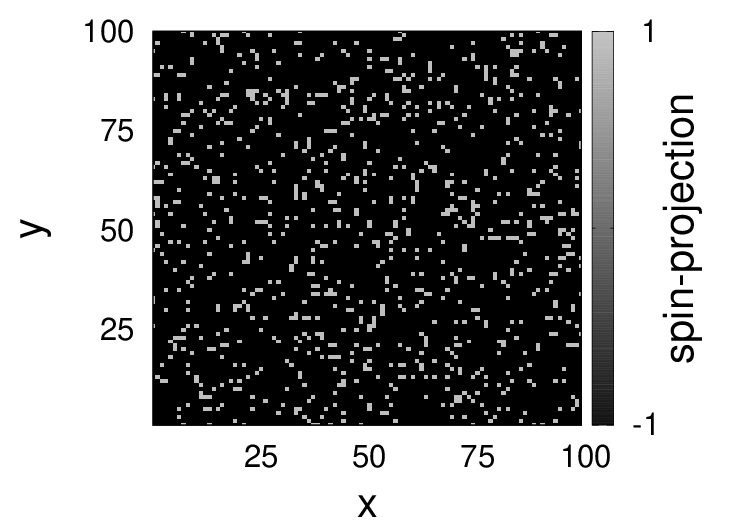}}
			
			\resizebox{6.0cm}{!}{\includegraphics[angle=0]{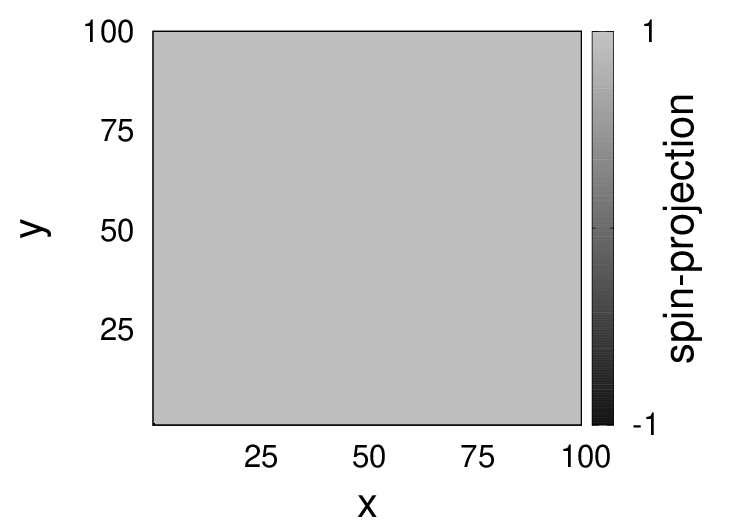}}\\
			
			\phantom{}
			\hspace{-0.00cm} $M_{t}(T,\sigma)=-0.189$ \hspace{3.00cm} $M_{m}(T,\sigma)=-0.814$ \hspace{2.75cm} $M_{b}(T,\sigma)=+1.000$\\
			
			\\

			$\mathbf{(b)\text{ } \sigma=0.20}$\\
			
			\resizebox{6.0cm}{!}{\includegraphics[angle=0]{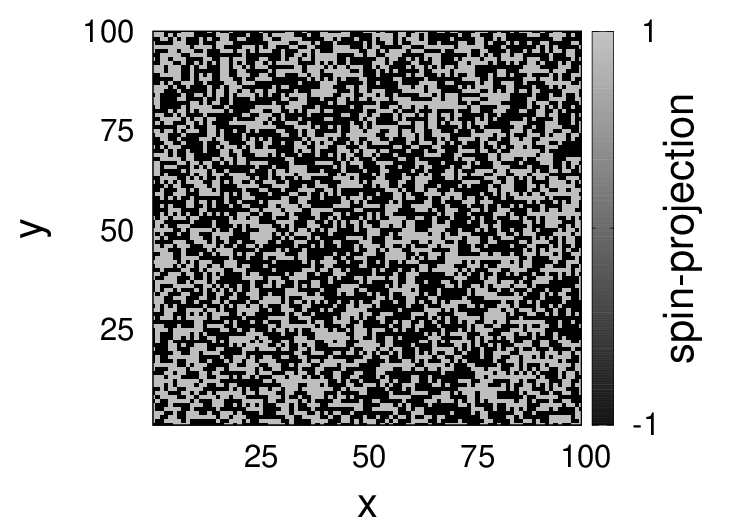}}
			
			\resizebox{6.0cm}{!}{\includegraphics[angle=0]{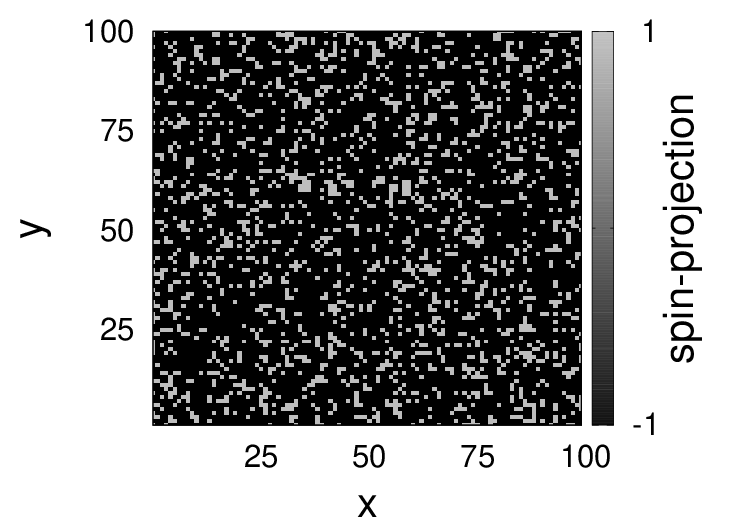}}
			
			\resizebox{6.0cm}{!}{\includegraphics[angle=0]{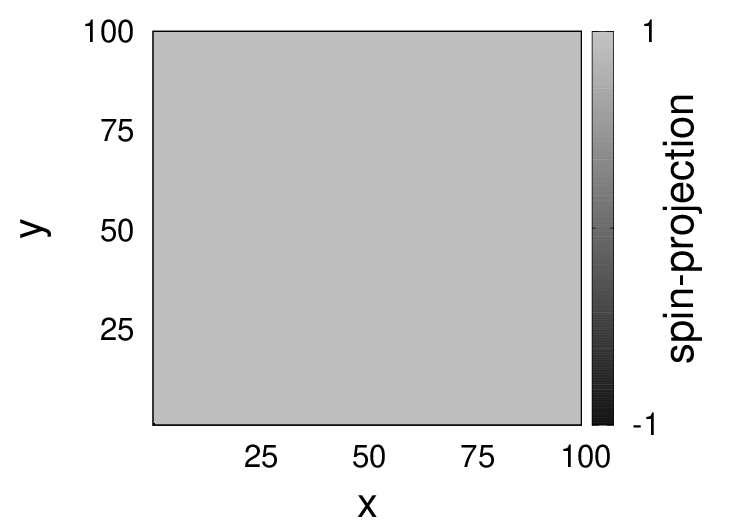}}\\
			
			\phantom{}
			\hspace{-0.00cm} $M_{t}(T,\sigma)=-0.091$ \hspace{3.00cm} $M_{m}(T,\sigma)=-0.615$ \hspace{2.75cm} $M_{b}(T,\sigma)=+1.000$\\
			
			\\
			
			$\mathbf{(c)\text{ } \sigma=0.76}$\\
			
			\resizebox{6.0cm}{!}{\includegraphics[angle=0]{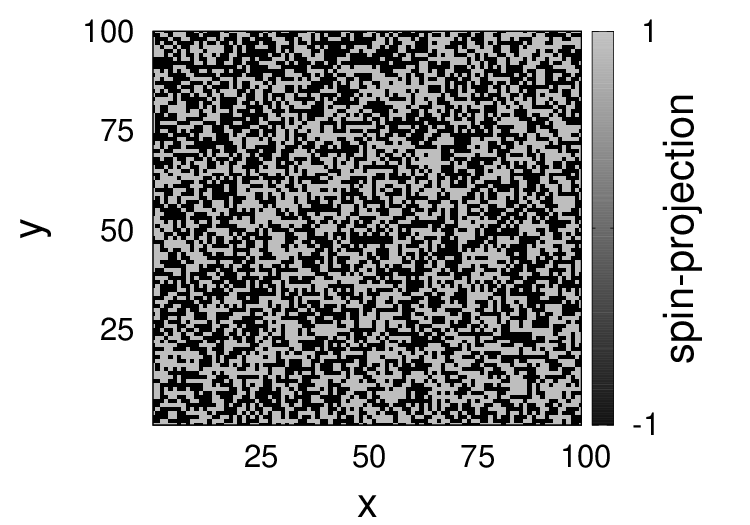}}
			
			\resizebox{6.0cm}{!}{\includegraphics[angle=0]{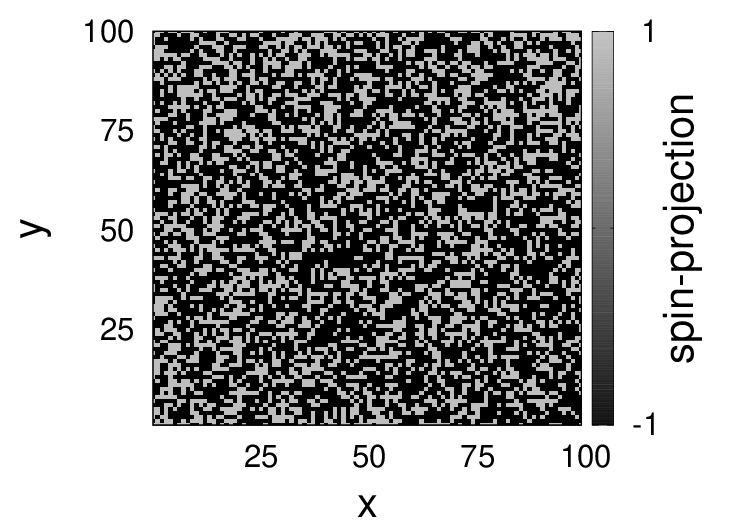}}
			
			\resizebox{6.0cm}{!}{\includegraphics[angle=0]{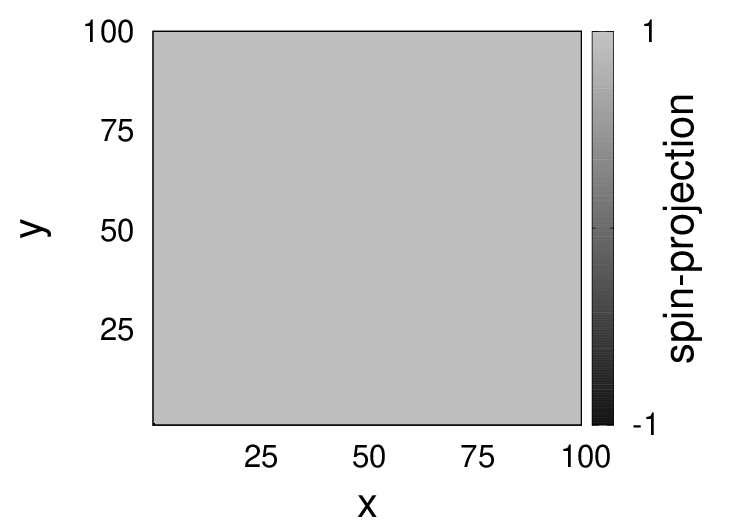}}\\
			
			\phantom{}
			\hspace{-0.00cm} $M_{t}(T,\sigma)=-0.022$ \hspace{3.00cm} $M_{m}(T,\sigma)=-0.240$ \hspace{2.75cm} $M_{b}(T,\sigma)=-1.000$\\
			
			\\
			
		\end{tabular}
		\caption{ \textbf{For AAB configuration}: Lattice morphologies of \textbf{top layer (at Left)}; \textbf{mid layer (at Middle)} and \textbf{bottom layer (at Right)} at $t=t_{morph}=10^{5}$ $MCS$ for $J_{AA}/J_{BB}=0.04$ and $J_{AB}/J_{BB}=-0.20$ and a few standard deviations of the external field. The magnetisations are rounded-off to three decimal places. The shift and vanishing of compensation in the respective cases (b) and (c) is due to the significant reduction of magnetic ordering in the top and middle A-layers.}
		\label{fig_morphology_AAB}
	\end{center}
\end{figure*}

To understand these arguments in the context of vanishing of compensation, let's study the lattice morphology (or, spin-density plots) at the \textit{zero-field compensation points} for both ABA and AAB configurations with $\sigma=\{0.00,0.20,0.76\}$ ($\sigma=0.00$ means absense of the external field). For the ABA configuration, as the external field is swept with $\sigma=0.20$, the surface A-layers lose significant magnetisation (i.e. decrease in magnetic ordering/increase in randomisation) at the $T_{comp}(\sigma=0)=0.502$ [Refer to Figure \ref{fig_morphology_ABA}]. Consequently we need to lower the temperature further to achieve the required magnetic ordering (i.e. magnetisation) in the surface A-layers, so that they cumulatively cancel out the magnetisation of the B-layer to produce a compensation point at a lower temperature value [Refer to Figure \ref{fig_mag_response_ABA}]. The surface A-layers almost behave identically. When the standard deviation of the external field is increased to $\sigma=0.76$, we can see (from the values of magnetisation beneath the spin-density plots), the magnetisation is very much reduced for the surface A-layers and conseqently, lowering of the temperature even further isn't able to generate enough magnetic ordering in the A-layers to cancel out the magnetization of the B-layer to create a compensation point leading to the field-driven vanishing of the compensation point in the steady state. The reduction of magnetic ordering leads to increase in the fluctuations of both, order parameter (equivalently, magnetisation) and cooperative energy per site. Following the similar argument we can readily understand the shift and vanishing of compensation in the AAB configuration when the standard deviation of the external field is increased for a fixed combination of coupling strengths [Refer to Figure \ref{fig_morphology_AAB}].

%%%%%%%%%%%%%%%%%%%%%%%%%%%%%%%%%%%%%%%%%%%%%%
\subsubsection{Magnetization versus time:}
\label{subsubsec_magvtime}
\begin{figure*}[!htb]
	\begin{center}
		\begin{tabular}{c}
			
			\resizebox{9.5cm}{!}{\includegraphics[angle=0]{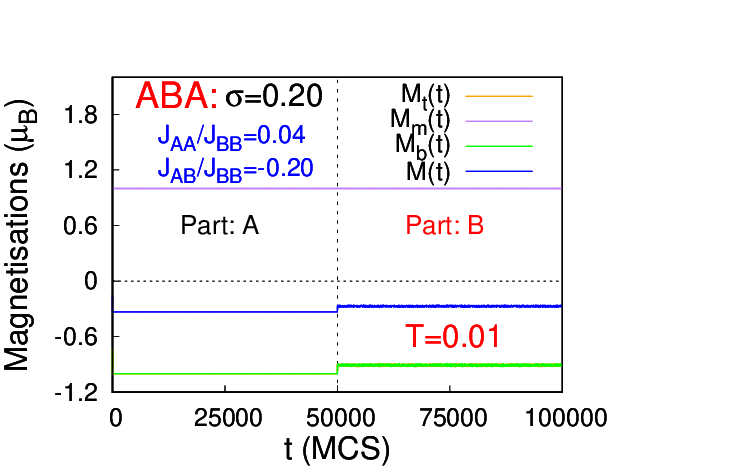}}
			\resizebox{9.5cm}{!}{\includegraphics[angle=0]{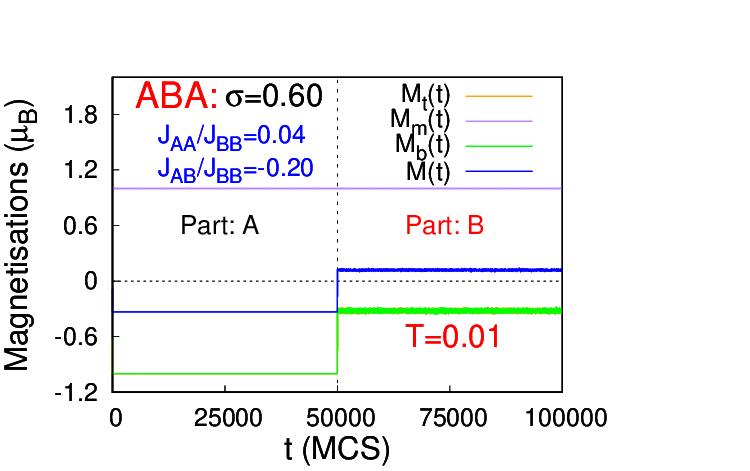}}\\
			
			\large {\textbf{(A) Top panel: $J_{AA}/J_{BB}=0.04$ and $J_{AB}/J_{BB}=-0.20$}}\\
			
			\resizebox{9.5cm}{!}{\includegraphics[angle=0]{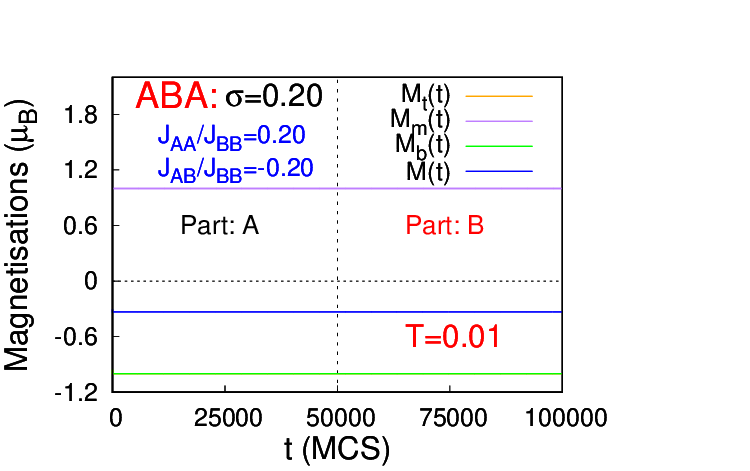}}
			\resizebox{9.5cm}{!}{\includegraphics[angle=0]{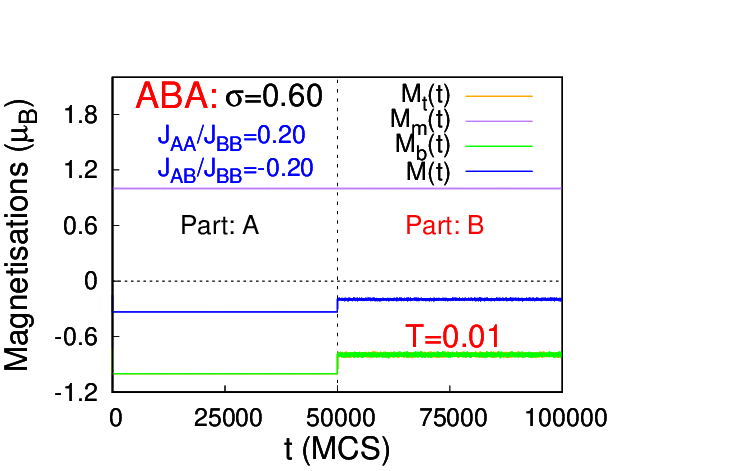}}\\
			
			\large {\textbf{(B) Middle panel: $J_{AA}/J_{BB}=0.20$ and $J_{AB}/J_{BB}=-0.20$}}\\
			
			\resizebox{9.5cm}{!}{\includegraphics[angle=0]{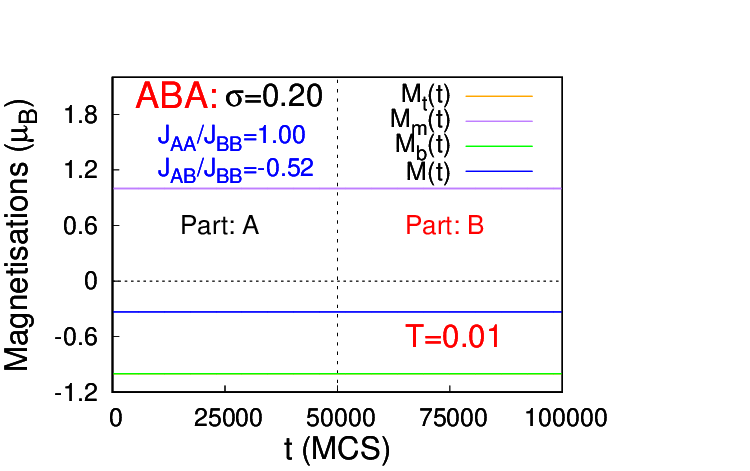}}
			\resizebox{9.5cm}{!}{\includegraphics[angle=0]{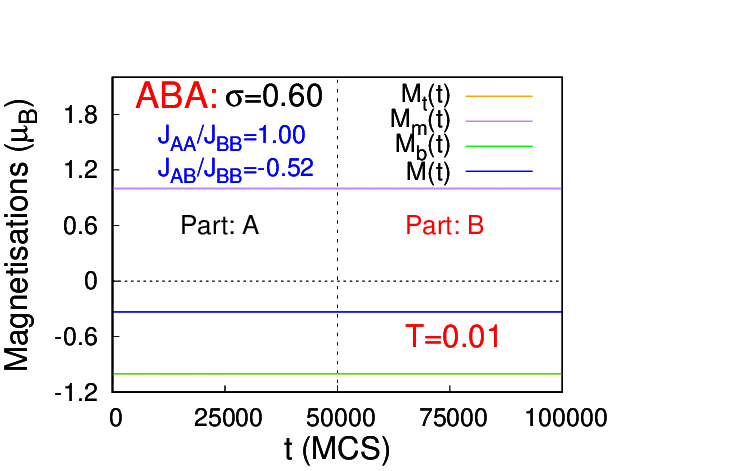}}\\		
			
			\large {\textbf{(C) Bottom panel: $J_{AA}/J_{BB}=1.00$ and $J_{AB}/J_{BB}=-0.52$}}\\
		\end{tabular}
		\caption{ (Colour Online) Plots of Magnetisations for square monolayers (sublattices) and total magnetisation of the bulk versus time in MCS, \textbf{for the \textit{ABA} configuration}, where $M_{t}(t)$: Magnetization of the top layer; $M_{m}(t)$: Magnetization of the mid layer; $M_{b}(t)$: Magnetization of the bottom layer are all functions of time,$t$, in units of MCS. In these figures, \textbf{Part: A} describes the equilibrium (Zero-field) and transient (Field: ON) behaviour whereas, \textbf{Part: B} describes the steady state behaviour (Field: ON). The magnetisation curves for the surface A-layers (orange and green) of ABA configuration overlap for the most of the times.}
		\label{fig_ABA_field_mag_time}
	\end{center}
\end{figure*}

\begin{figure*}[!htb]
	\begin{center}
		\begin{tabular}{c}
			
			\resizebox{8.5cm}{!}{\includegraphics[angle=0]{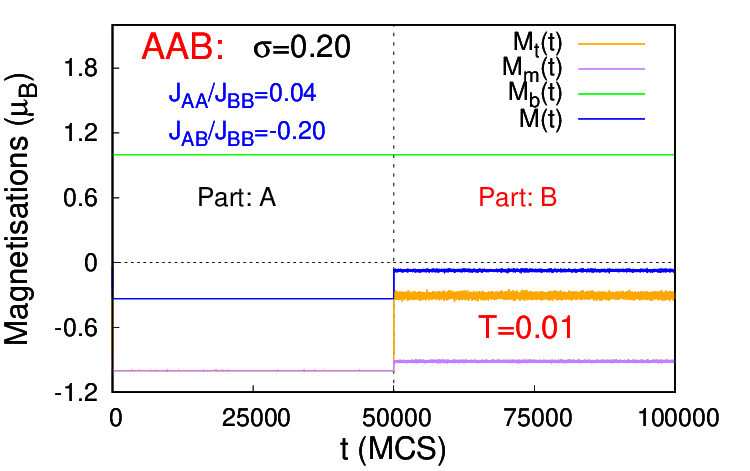}}
			\resizebox{8.5cm}{!}{\includegraphics[angle=0]{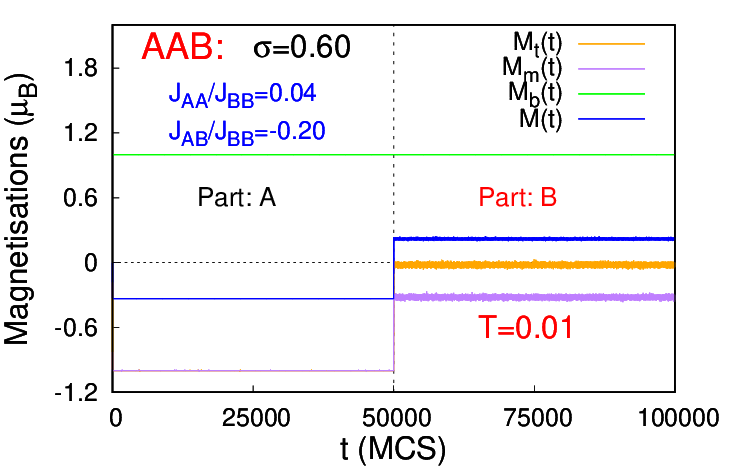}}\\
			
			\large {\textbf{(A) Top panel: $J_{AA}/J_{BB}=0.04$ and $J_{AB}/J_{BB}=-0.20$}}\\
			
			\resizebox{8.5cm}{!}{\includegraphics[angle=0]{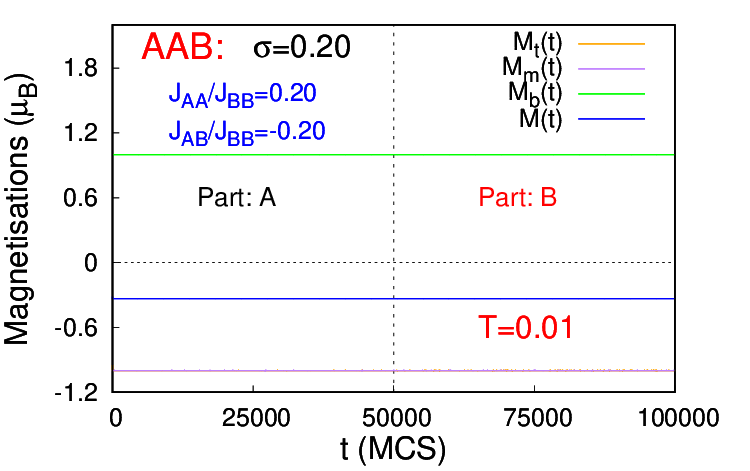}}
			\resizebox{8.5cm}{!}{\includegraphics[angle=0]{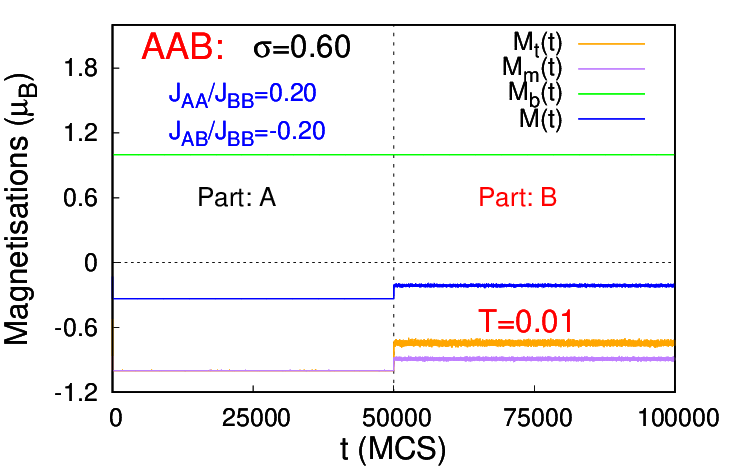}}\\
			
			\large {\textbf{(B) Middle panel: $J_{AA}/J_{BB}=0.20$ and $J_{AB}/J_{BB}=-0.20$}}\\
			
			\resizebox{8.5cm}{!}{\includegraphics[angle=0]{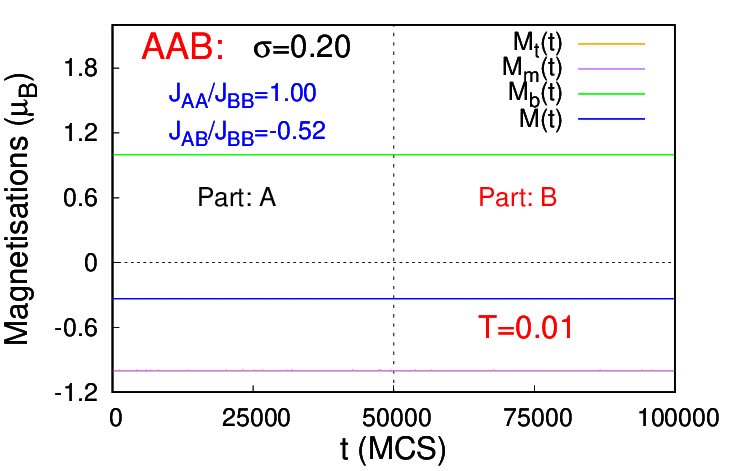}}
			\resizebox{8.5cm}{!}{\includegraphics[angle=0]{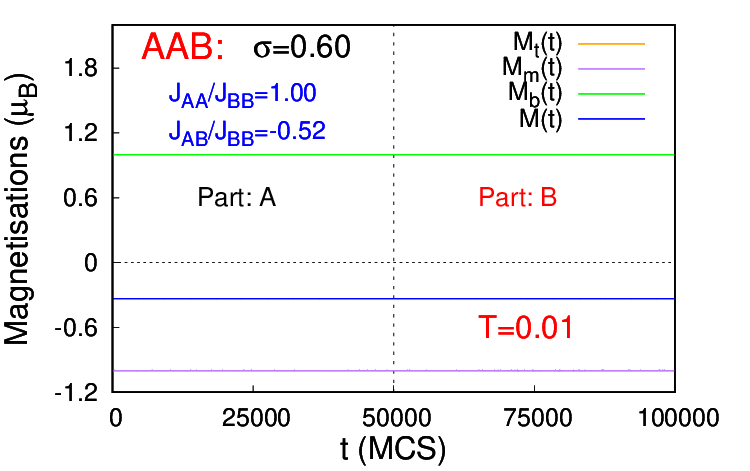}}\\			
			
			\large {\textbf{(C) Bottom panel: $J_{AA}/J_{BB}=1.00$ and $J_{AB}/J_{BB}=-0.52$}}\\
		\end{tabular}
		\caption{ (Colour Online) Plots of Magnetisations for square monolayers (sublattices) and total magnetisation of the bulk versus time in MCS, \textbf{for the \textit{AAB} configuration}, where $M_{t}(t)$: Magnetization of the top layer; $M_{m}(t)$: Magnetization of the mid layer; $M_{b}(t)$: Magnetization of the bottom layer are all functions of time,$t$, in units of MCS. In these figures, \textbf{Part: A} describes the equilibrium (Zero-field) and transient (Field: ON) behaviour whereas, \textbf{Part: B} describes the steady state behaviour (Field: ON).}
		\label{fig_AAB_field_mag_time}
	\end{center}
\end{figure*}

%%%%%%%%%%%%%%%%%%%%%%%%%%%%%%%%%%%%%%%%%%%%%%
\indent The section is devoted to the behaviour of sublattice magnetisations with time as the field is switched ON, at a suitably very low temperature $T=0.01$. We have chosen three distinct combinations of the coupling ratios, $J_{AA}/J_{BB}$ and $J_{AB}/J_{BB}$, as (0.04,-0.20), (0.20,-0.20), (1.00,-0.52) for the two values of the standard deviation, $\sigma=$\{0.20,0.60\} of the external Gaussian random field for both the ABA and AAB configurations [Figures \ref{fig_ABA_field_mag_time} and \ref{fig_AAB_field_mag_time}]. At $T=0.01$, till $t= 5\times 10^{4}$ MCS, the Hamiltonian only has the cooperative part and the sublattice and total magnetisation of the system remains in equilibrium. Just after $t=t_{0}=5 \times 10^{4}$ MCS, the external field is switched ON and the sublayer and total magnetisations start to react. As in the case with uniform random field \cite{Chandra5}, both the systems, ABA and AAB, reach the steady state very quickly. Conclusive features are unraveled in these cases.

\textbf{For the ABA configuration}, we see \textbf{in the top panel: Figure \ref{fig_ABA_field_mag_time}(A)}, with $J_{AA}/J_{BB}=0.04$ and $J_{AB}/J_{BB}=-0.20$, both the surface A-layers react magnetically for both the standard deviations of the external Gaussian field. The reason is, the per site cooperative energy of the A-layers is comparable to the spin-field energy. But the mid B-layer with dominant in-plane coupling, remains in its equilibrium magnetic state. So it is evident that the reduction (or, destruction) of magnetic order in the sublayers is the result of the competition between the cooperative and spin-field energies and spin-field energies dominating the cooperative part in the relevant cases. The similar inference is valid for \textbf{the middle panel: Figure \ref{fig_ABA_field_mag_time} (B)}, with the $\sigma=0.60$, where per site spin-field energies are comparable to the cooperative part of the Hamiltonian for the surface A-layers. In \textbf{the bottom panel: Figure \ref{fig_ABA_field_mag_time} (C)}, with the $\sigma=0.20$ and $\sigma=0.60$, all the three sublayers don't react. Here, the cooperative part of the Hamiltonian for the surface A-layers dominates the steady state per site spin-field energies. The magnetization curves for the top and bottom layers overlap for most of the times as they have identical interacting magnetic neighborhood. In \textbf{Figure \ref{fig_ABA_field_mag_time}(A)} with $J_{AA}/J_{BB}=0.04$, $J_{AB}/J_{BB}=-0.20$ and $\sigma=0.60$ , we see the reduction in magnetization of the surface layers in presence of the external field leads to the vanishing of compensation even in the lowest possible temperature in simulation. Even in the lowest simulational temperature, the cumulative steady state value of magnetization of the surface A-layers becomes smaller than even the steady state value of magnetization of the middle B-layer (that is why the total magnetisation remains positive, same signature as the B-layer, in the presence of the field).  Even in the lowest temperature, $$|\langle M_{t}(T) + M_{b}(T) \rangle| < |\langle M_{m}(T)\rangle|$$ which causes compensation to disappear in the presence of the field. This is the reason behind all the instances with field-driven vanishing of compensation.

\textbf{For the AAB configuration}, the influence of the bottom B-layer is limited to the middle A-layer (because of the nearest neighbour Ising interactions). So the top A-layer gets much more affected by the external field than the middle A-layer. In Figure \ref{fig_AAB_field_mag_time}, we can understand it by simply noticing the orange line for the magnetisation of the top A-layer. \textbf{In the top panel: Figure \ref{fig_AAB_field_mag_time}(A)}, with  $J_{AA}/J_{BB}=0.04$, $J_{AB}/J_{BB}=-0.20$ and $\sigma={0.60,1.00}$, the spin-field energies per site dominates the cooperative energy per site of the A-layers. Consequuently, the A layers lose much of the magnetic ordering at even the lowest temperature, the extent of randomisation is much more prominent for the top A-layer and the top A-layer is almost completely randomised when $\sigma=0.60$. So the combined magnetisation of A-layers is \textit{not enough to nullify the magnetisation of the bottom B-layer}, leading to vanishing of compensation. \textbf{In Figures \ref{fig_AAB_field_mag_time}(B) and (C)}, we can explain the behaviour in light of the discussions above. A few more instances of both the configurations for $\sigma=1.00$ are presented in Appendix \ref{app_high_sigma} to supplement this discussion for a better understanding. Such behaviour is qualitatively similar to what we have seen with uniform random field \cite{Chandra5}.

This is another example of \textbf{dynamic field-driven vanishing of compensation} in the Ising spin-1/2 trilayers, driven by Gaussian random external field with spatiotemporal variation. The bottom panel with $J_{AA}/J_{BB}=1.00$ and $J_{AB}/J_{BB}=-0.52$ for both the configurations supports that vanishing of compensation is a result of the competition between the cooperative and spin-field energies. 

%%%%%%%%%%%%%%%%%%%%%%%%%%%%%%%%%%%%%%%%%%%%%

\subsection{Phase Diagram and Scaling}
%%%%%%%%%%%%%%%%%%%%%%%%%%%%%%%%%%%%%%%%%%%%

\begin{figure*}[!htb]
	\begin{center}
		\begin{tabular}{c}
			\textbf{(a)}
			\resizebox{8cm}{!}{\includegraphics[angle=0]{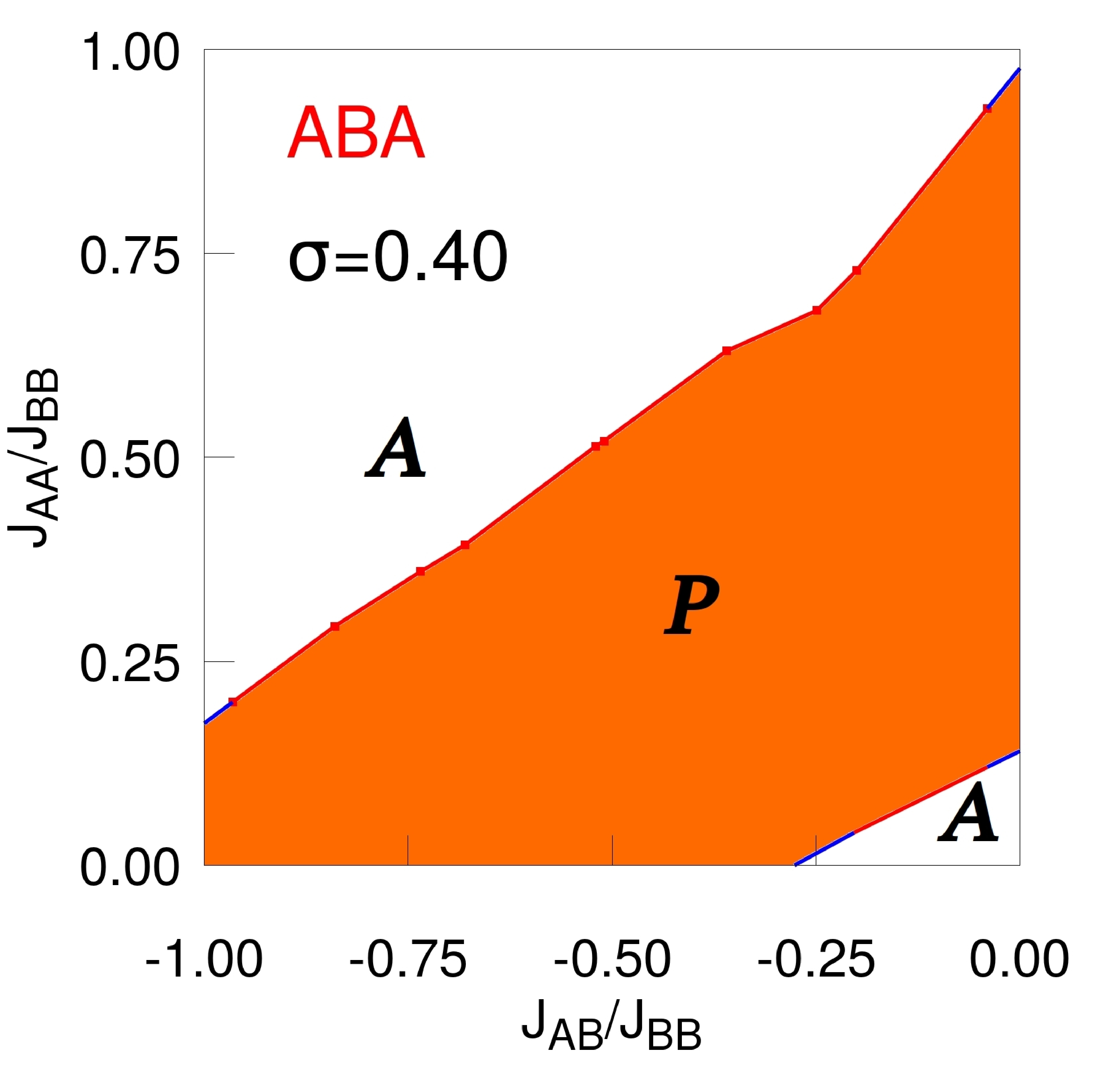}}
			\hfill
			\textbf{(b)}
			\resizebox{8cm}{!}{\includegraphics[angle=0]{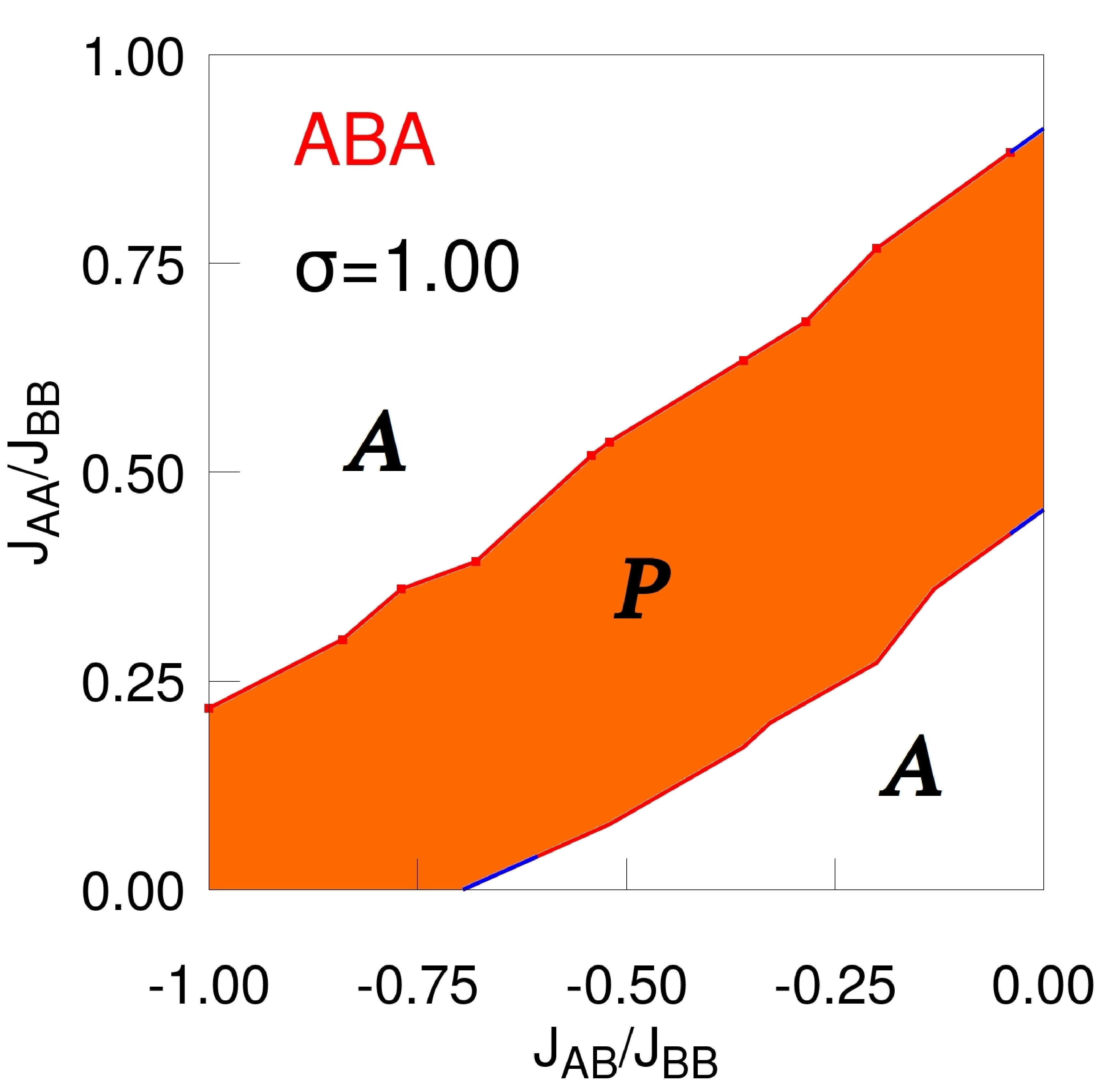}}\\
			
			\textbf{(c)}
			\resizebox{8cm}{!}{\includegraphics[angle=0]{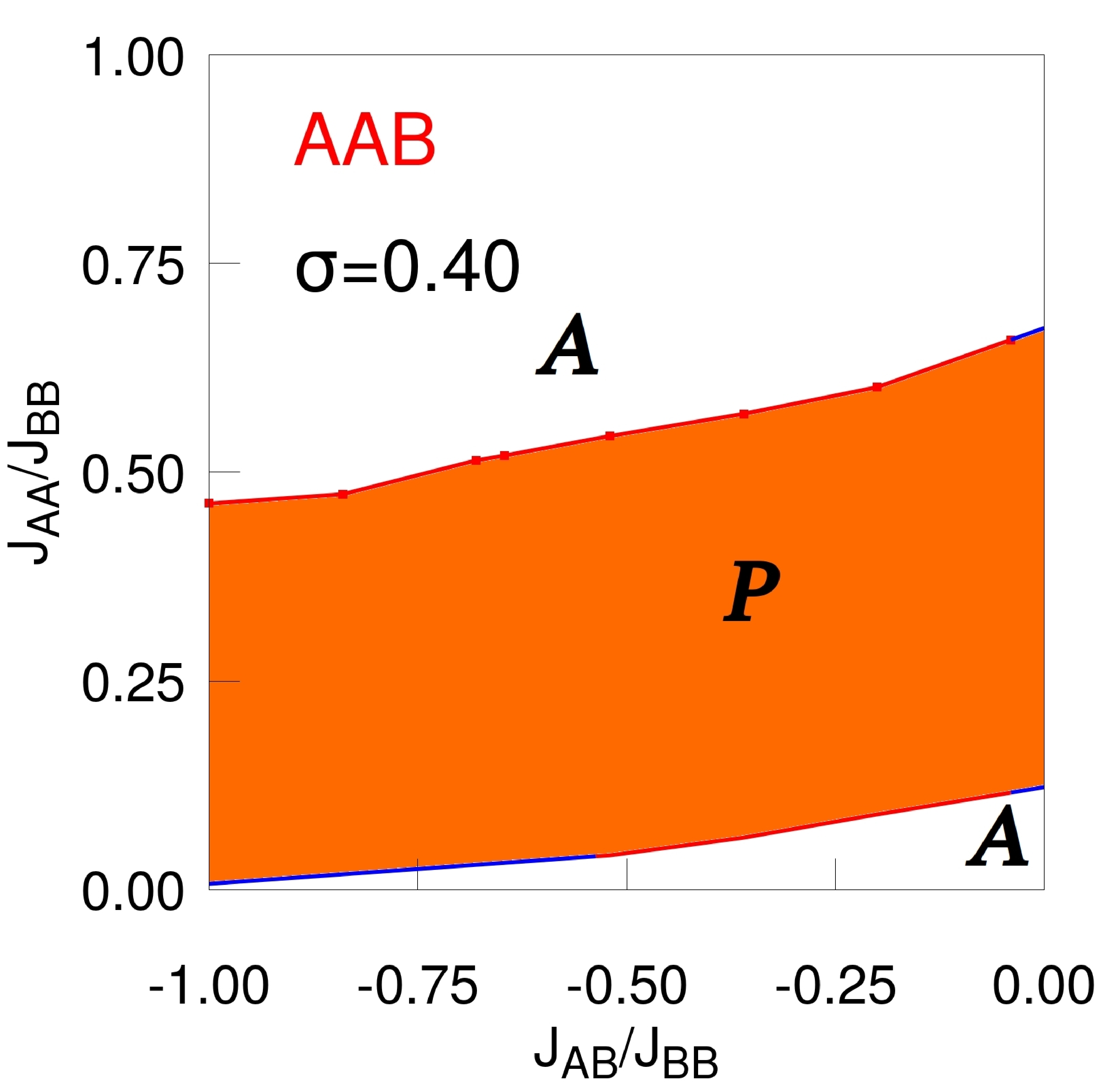}}
			\hfill
			\textbf{(d)}
			\resizebox{8cm}{!}{\includegraphics[angle=0]{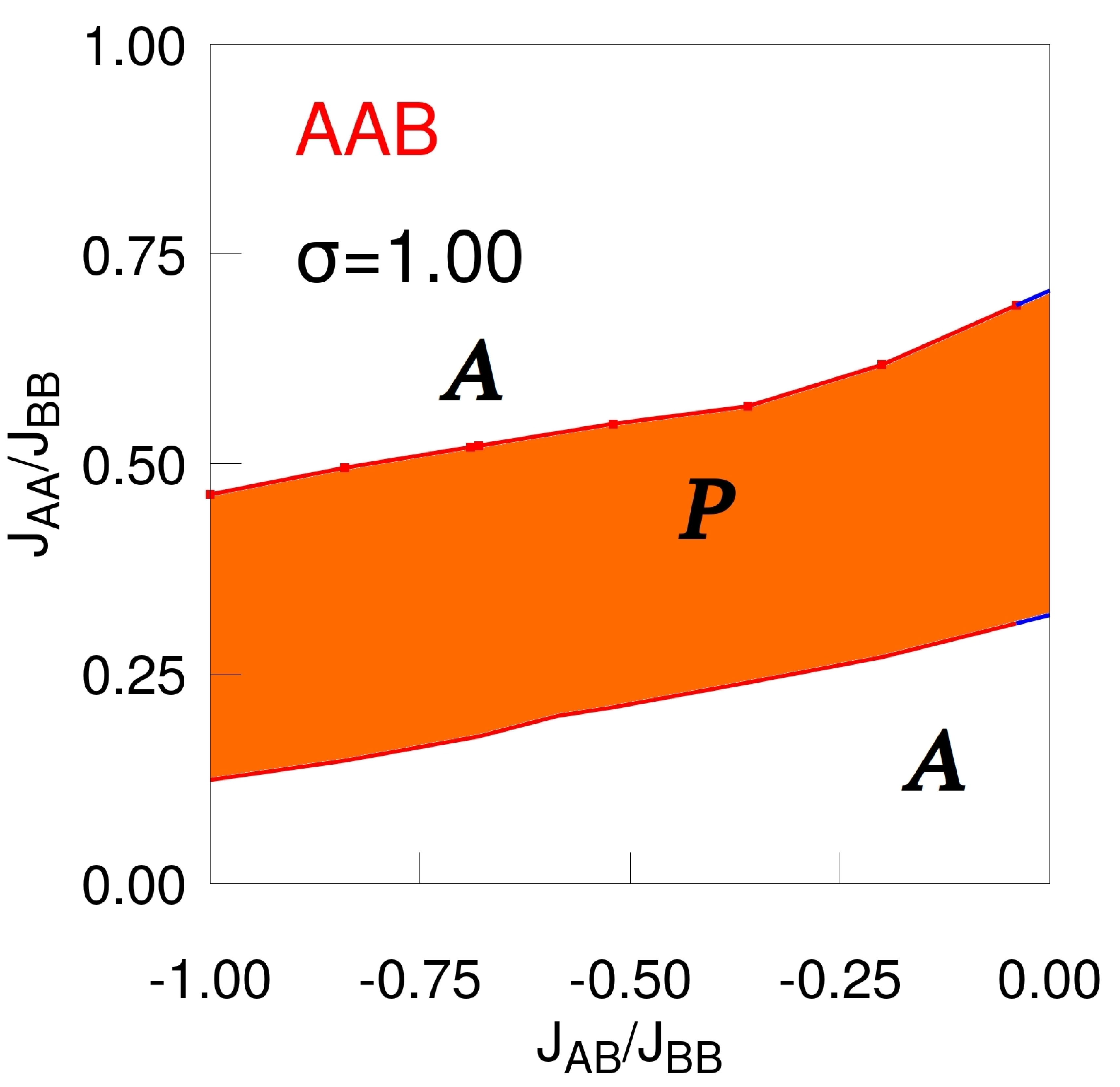}}
			
		\end{tabular}
		\caption{ (Colour Online) Phase diagram for the: ABA trilayered ferrimagnetic system when: (a) $\sigma=0.40$; (b) $\sigma=1.00$ and AAB trilayered ferrimagnetic system when: (c) $\sigma=0.40$; (d) $\sigma=1.00$, in presence of the uniform random external magnetic field. A: Compensation is absent; P: Compensation is present. With an increase in the standard deviation of the external field, the magnitude of the area of the no-compensation island have grown. The blue segment of the phase separation curves are obtained via linear extrapolation. All these plots are obtained for a system of $3\times100\times100$ sites. Where the errorbars are not visible, they are smaller than the point markers.}
		\label{fig_phasecurve_2d}
	\end{center}
\end{figure*}

\begin{figure*}[!htb]
	\begin{center}
		\begin{tabular}{c}
			\textbf{(a)}
			\resizebox{8.5cm}{!}{\includegraphics[angle=-90]{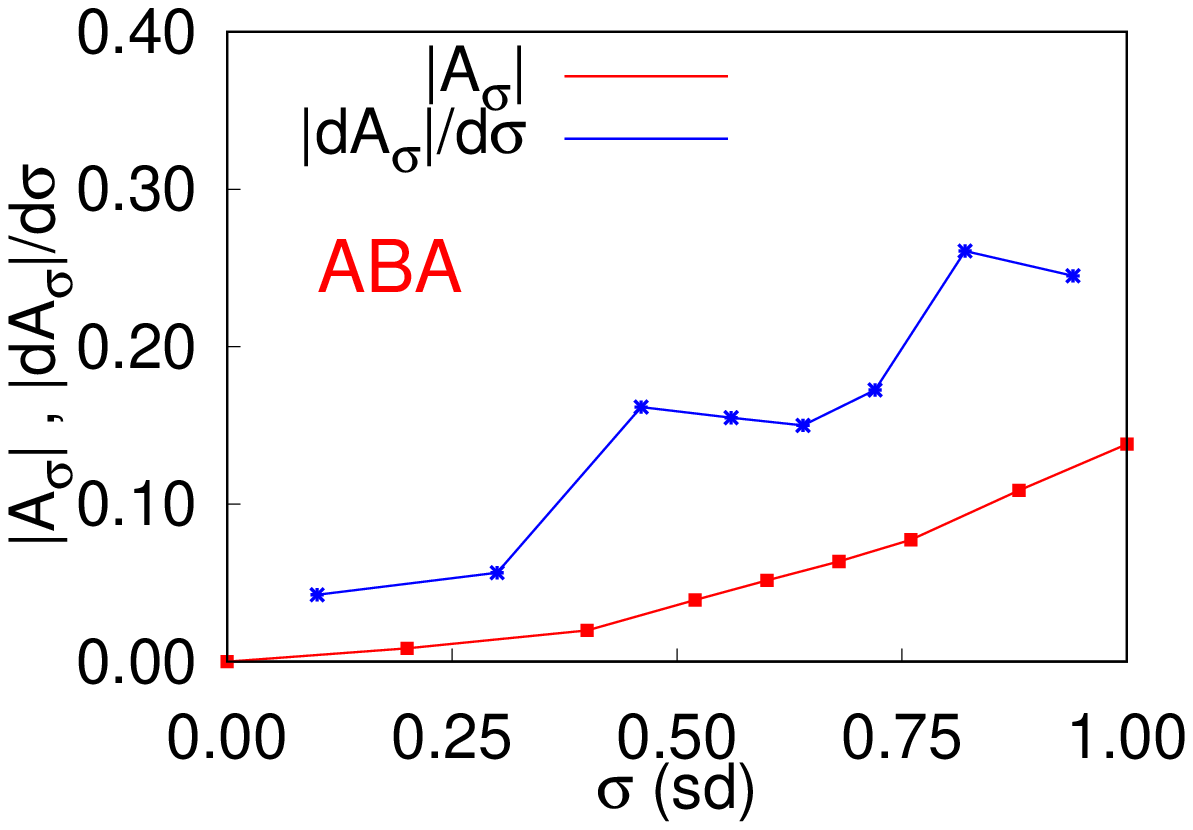}}
			\hfill
			\textbf{(b)}
			\resizebox{8.5cm}{!}{\includegraphics[angle=-90]{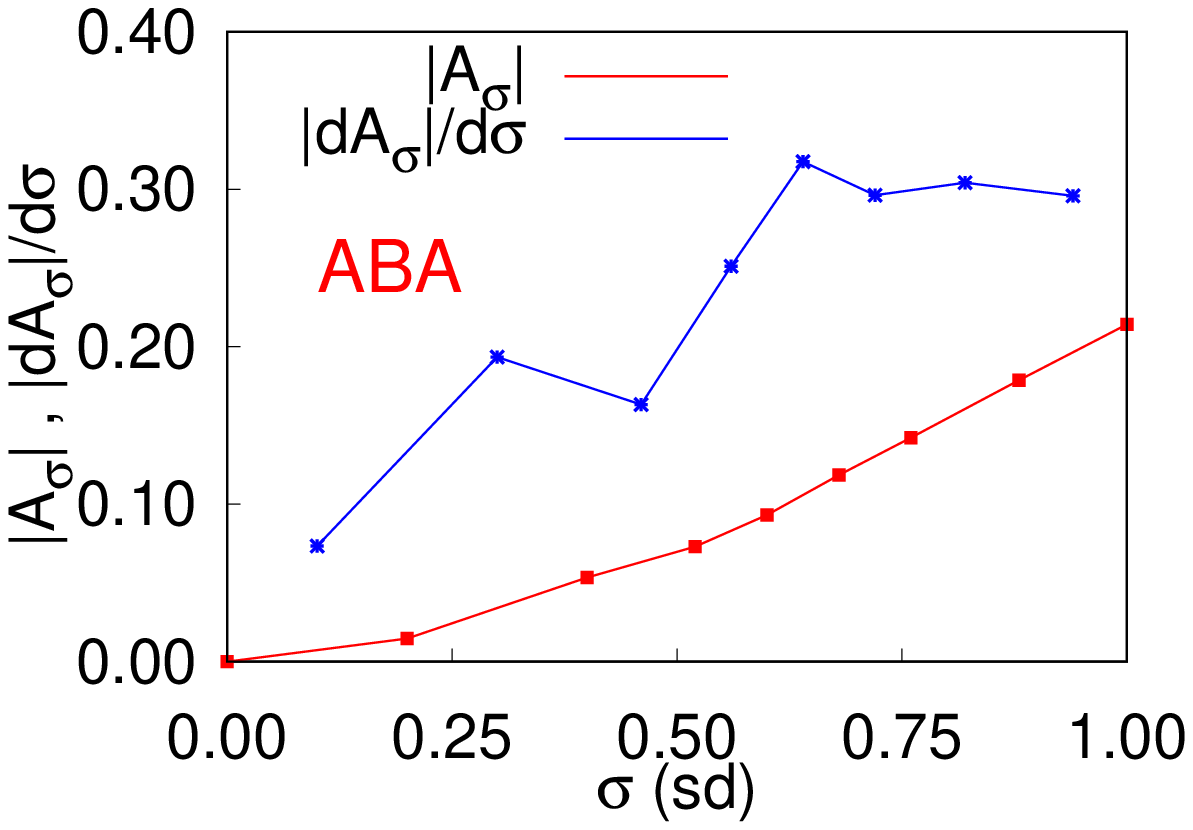}}
			
		\end{tabular}
		\caption{ (Colour Online) Plots of: Magnitude of the area of the no-compensation islands versus standard deviation of the field (in RED) and the rate of increase in the magnitude of the area of the no-compensation islands versus standard deviation of the field (in BLUE) for (a) ABA and (b) AAB configurations for a system of $3\times100\times100$ sites.}
		\label{fig_area_slope_field}
	\end{center}
\end{figure*}

\indent The phase diagrams in Figure \ref{fig_phasecurve_2d} depict the effects of a Gaussian random external field on the Hamiltonian parameter space for both the trilayered ferrimagnetic systems. Compensation temperature merges with the critical temperature for higher values of $J_{AA}/J_{BB}$ when $|J_{AB}/J_{BB}|$ is fixed or vice-versa, just like in the zero-field case. In Figure \ref{fig_phasecurve_2d}, the phase diagrams are drawn following the zero-field cases \cite{Diaz1,Diaz2,Chandra3} where, compensation is present (marked by P) within the orange areas and absence of compensation is marked by white areas (marked by A). The presence of the external Gaussian field [From $\sigma$=0.2, onwards] creates the enclave or islands with no-compensation within the phase area where parameters support compensation. These closed areas or No-Compensation Islands (NCI) grow as the randomness (equivalently, standard deviation) of the external field increases, similar to the case with uniform random external field \cite{Chandra5}. In Figure \ref{fig_area_slope_field}, we present the plots of absolute area and rate of increase of absolute area versus the standard deviation of the applied field. Linear interpolation/extrapolation is employed to obtain the closed curve for the boundary of the NCI, and the fractional area is obtained by Monte Carlo Integration \cite{Krauth}. Central difference formula is employed to find out the rate of increase of the area of NCIs to determine the nature of curve of the area of NCI versus standard deviation of the external field.

To analyse the nature of the curves of absolute area versus the standard deviation, we note, the curve (in RED), comes out to be a mixture of \textit{superlinear and sublinear} in nature, \textbf{for the ABA configuration}. From the BLUE curve of slope versus standard deviation of the field in Figure \ref{fig_area_slope_field}(a), in the vicinity of $\sigma=0.60$  we find the curve is sublinear (almost linear). Again at and after $\sigma=0.88$, the area changes in a prominent sublinear manner. At the low random fields around $\sigma=0.20$ and around moderately high randomness around $\sigma=0.70$, the nature is superlinear. \textbf{For the AAB configuration}, the area of NCIs increase superlinearly on average till $\sigma=0.64$. After that the behaviour is almost linear. Now the scaling between the magnitude of the area of NCIs and the standard deviation of the field can be performed by the following scaling function \cite{Chandra5}:
\begin{equation}
\label{eq_scaling}
	f(A(\sigma), \sigma)= \sigma^{-b}A(\sigma) 
\end{equation}
The scaling exponents come out to be: for ABA: $b_{ABA}=1.913\pm0.137$ and for AAB: $b_{AAB}=1.625\pm0.066$. A faithful estimate of error in the values of the exponents is obtained by the standard deviation among all the sets of data. 
%%%%%%%%%%%%%%%%%%%%%%%%%%%%%%%%%%%%%%%%%%%%%
%%%%%%%%%%%%%%%%%%%%%%%%%%%%%%%%%%%%%%%%%%%%%
\section{Summary and Conclusion}
\label{sec_summary}
%%%%%%%%%%%%%%%%%%%%%%%%%%%%%%%%%%%%%%%%%%%%%

In the Ising model, the coupling constants are traditionally taken to be translationally invariant. Along with that, competing ferromagnetic and antiferromagnetic interactions in the systems of Figure \ref{fig_lattice_structure} throw up exciting bulk behaviour e.g. Compensation. The equilibrium studies on these systems \cite{Diaz1,Diaz2,Chandra1,Chandra2} have shown us the prevalent complexity in deriving conditions for the existence of compensation. Now, in the current article, a Metropolis Monte Carlo study has been performed on the magnetic and thermodynamic responses of the systems of Figure \ref{fig_lattice_structure} under the influence of a Gaussian Random external field with spatio-temporal variations.

It is time to discuss the implications of the current work. From \cite{Chandra5}, we have a fair idea about how such systems react under the influence of uniform random magnetic field. From Section \ref{subsubsec_magvtemp} we find that the magnetic response is qualitatively similar to the reponses under the uniform random field. The compensation and critical temperatures shift towards the low temperature ends and even results in the vanishing of compensation in proper cases as we increase the standard deviation of the external Gaussian random magnetic field. Similar inference can be drawn from the thermodynamic behaviour of the suitably defined fluctuations of magnetization and cooperative energy [Refer to Section \ref{subsubsec_flucvtemp}]. The effect of the time-dependent part of the Hamiltonian is established in Section \ref{subsubsec_magvtime}. We observe that the systems react very quickly after switching ON the external field and the dynamics is governed by the competition between spin-field energies and cooperative energies. Thus the Gaussian random field-driven vanishing of compensation, observed in this work, is also a dynamic phenomenon like it was in \cite{Chandra5} with a uniform random external field. The phase diagrams in Figure \ref{fig_phasecurve_2d}, for both the ABA and AAB configurations, have similar kind of appearance with No-Compensation Islands engraved within the phase area with compensation. As we investigate the plot of the magnitude of the area of NO-Compensation Islands versus the standard deviation of the external field and find out the scaling exponent according to the Equation \ref{eq_scaling}, we find the responses are qualitatively similar for both the continuous field distributions: Uniform and Gaussian. A quick comparison follows: $b_{ABA}^{Uniform} = 1.958 \pm 0.122$ and
$b_{ABA}^{Gaussian} = 1.913 \pm 0.137$ (\textbf{For ABA}) and $b_{AAB}^{Uniform} = 1.783 \pm 0.118$ and
$b_{AAB}^{Gaussian} = 1.625 \pm 0.066$ (\textbf{For AAB}), where data for the uniform random field are taken from \cite{Chandra5}. So, there exists a very good agreement for the scaling exponents as they overlap within the statistical interval of one-another. So the dynamic response of the trilayered Ising spin-1/2 square ferrimagnnetic systems, in this article, are quite similar under the influence of Gaussian random external magnetic field with spatiotemporal variations listed in Section \ref{app_char_field} to the Uniform random external field \cite{Chandra5}. The exact nature of the external continuous field distribution, for these two types of distributions, does not show a distinguished effect on the qualitative and quantitative features of such systems. But the results also pose a difficulty for technological applications. It is difficult to create a source of purely static magnetic field, as some kind of ripple may exist. That ripple or time dependent part, following uniform or Gaussian distribution with characteristics described here or in \cite{Chandra5}, may shift the compensation and critical temperatures from designated values.

Still we have unanswered questions. For example, how would the system behave under the influence of an external magnetic field following a Simpson distribution \cite{Wentzel,Alder}, with spatio-temporal variation of similar kind of the present work? The results would definitely help us comment strongly on the behaviour of the scaling exponent, $b$, under a wide variety of continuous field distributions for both the ABA and AAB type of trilayered stackings. The responses under the discrete distributions are also yet to be reported. these are planned for the future. In real magnetic systems, impurities, compositional disorder, lattice dislocations etc. modify the Hamiltonian to a translationally non-invariant kind. Such a complexity, with Ising mechanics, may be described by a dynamic Hamiltonian such as Equation \ref{eq_Hamiltonian} with a time dependent part, where the external field is characterized by a probability distribution.

%%%%%%%%%%%%%%%%%%%%%%%%%%%%%%%%%%%%%%%%%%%%
%%%%%%%%%%%%%%%%%%%%%%%%%%%%%%%%%%%%%%%%%%%%
\section*{Conflicts of interest}
%%%%%%%%%%%%%%%%%%%%%%%%%%%%%%%%%%%%%%%%%%%%
There are no conflicts of interest to declare.

%%%%%%%%%%%%%%%%%%%%%%%%%%%%%%%%%%%%%%%%%%%%

%%%%%%%%%%%%%%%%%%%%%%%%%%%%%%%%%%%%%%%%%%%%
\section*{Data availability statement}
%%%%%%%%%%%%%%%%%%%%%%%%%%%%%%%%%%%%%%%%%%%%
The data that support the findings of this study are available from the author upon reasonable request.

%%%%%%%%%%%%%%%%%%%%%%%%%%%%%%%%%%%%%%%%%%%%

%%%%%%%%%%%%%%%%%%%%%%%%%%%%%%%%%%%%%%%%%%%%
\section*{Acknowledgements}
%%%%%%%%%%%%%%%%%%%%%%%%%%%%%%%%%%%%%%%%%%%%
The author acknowledges the financial assistance from the University Grants Commission, India in the form of Research Fellowship and extends his thanks to Dr. Tamaghna Maitra for feedback and technical assistance.
%%%%%%%%%%%%%%%%%%%%%%%%%%%%%%%%%%%%%%%%%%%%
\newpage

\flushleft 
\textbf{\Large{Appendix}}

\appendix
%%%%%%%%%%%%%%%%%%%%%%%%%%%%%%%%%%%%%%%%%%%%%
\section{Characteristics of the External Magnetic Field}
\label{app_char_field}
%%%%%%%%%%%%%%%%%%%%%%%%%%%%%%%%%%%%%%%%%%%%%
\indent The local, Gaussian random external magnetic field values [of Equation \ref{eq_Hamiltonian}] at any site, at a time instant follow a Gaussian probability distribution. Box-Muller algorithm \cite{Box} is used to get such a distribution $G_{0}$ of standard deviation, $\sigma$ and zero mean:

\begin{equation}
G_{0}=\sigma \sqrt{-2 ln(U_{1})} cos(2\pi U_{2}) 
\end{equation}

Here $U_{1}$ and $U_{2}$ are two uniform random distributions between $\left[0,1\right]$ .

\indent A few additional characteristics are also added to the external field distribution \cite{Chandra5}:
\begin{itemize}
	\item[(a)] At different lattice sites, the values of the external field are uncorrelated at any time instant. Again at a lattice site, the values of the external field are uncorrelated for different time instants. So :  $h_{m}(t)h_{n}(t^{\prime})=a(t)\text{ }\delta_{mn}\text{ }\delta (t-t^{\prime})$, where $m,n$ are two different lattice sites and $t,t^{\prime}$ are two different time instants. 
	
	\item[(b)] The following conditions are also satisfied:
	\begin{itemize}
		\item[(i)] After the field is switched ON, the bulk average of the Gaussian field at a time instant $t$, is zero: $$\sum_{m} h_{m}(t) =0$$.\\
		So, $$ \sum_{m,n} h_{m}(t)h_{n}(t)\delta_{mn} =3L^{2}\sigma^{2}$$.
		
		\item[(ii)] At the $m$-th site, the temporal mean of the local field over the exposure interval, $\delta$, is \textit{zero}: $\langle h_{m}(t)\rangle=\dfrac{1}{\delta}\int_{t_{0}}^{t_{0}+\delta}  h_{m}(t)dt=0$ .
		
	\end{itemize}	
	
\end{itemize}

\indent At a few randomly chosen time instants within the exposure interval, the implementation of the desired field distribution is checked by the Cumulative Distribution Function (CDF) \cite{Deisenroth}, the Kernel Density Estimate (KDE) \cite{Rosenblatt-Parzen} and the Histogram. 
%%%%%%%%%%%%%%%%%%%%%%%%%%%%%%%%%%%%%%%%%%%%%
%%%%%%%%%%%%%%%%%%%%%%%%%%%%%%%%%%%%%%%%%%%%%
\section{On Compensation point and size of the system}
\label{app_tcomp_size_field}
%%%%%%%%%%%%%%%%%%%%%%%%%%%%%%%%%%%%%%%%%%%%%

\begin{figure*}[!htb]
	\begin{center}
		\begin{tabular}{c}
			
			\resizebox{8.5cm}{!}{\includegraphics[angle=0]{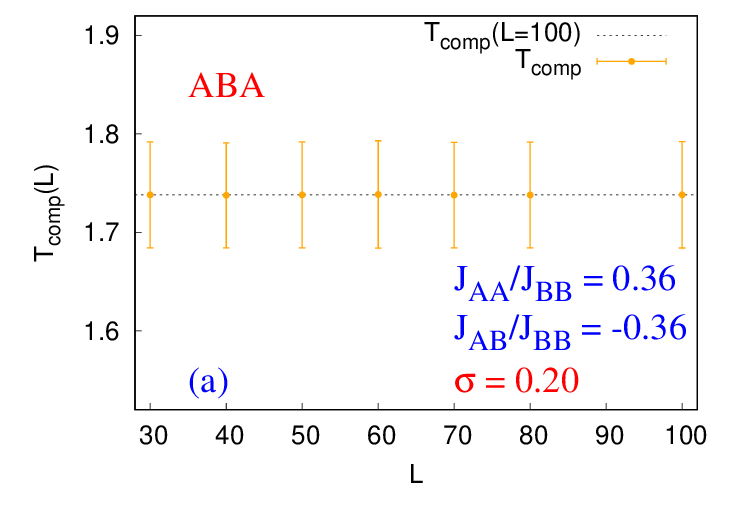}}
			
			\resizebox{8.5cm}{!}{\includegraphics[angle=0]{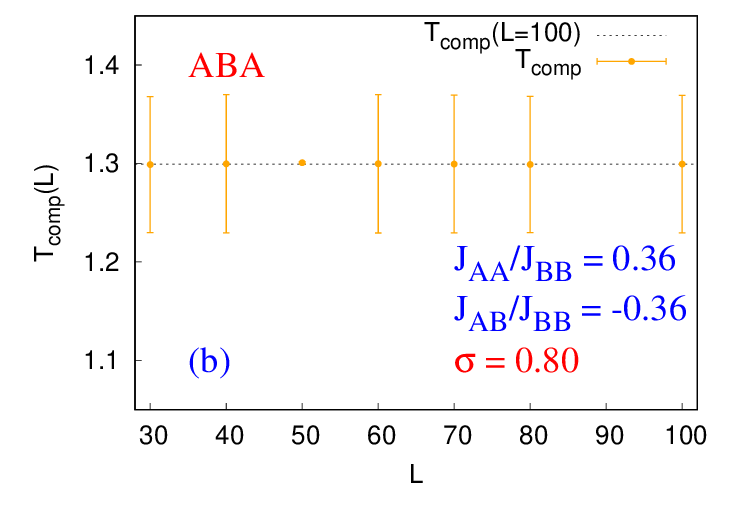}}\\
			
			\resizebox{8.5cm}{!}{\includegraphics[angle=0]{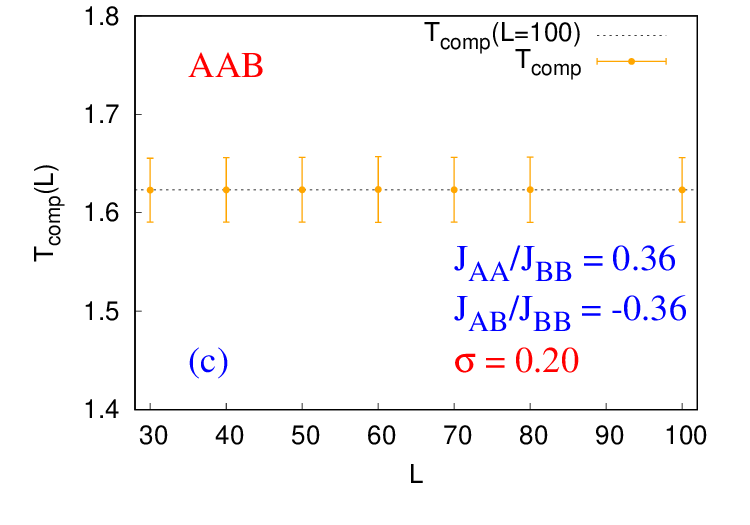}}
			
			\resizebox{8.5cm}{!}{\includegraphics[angle=0]{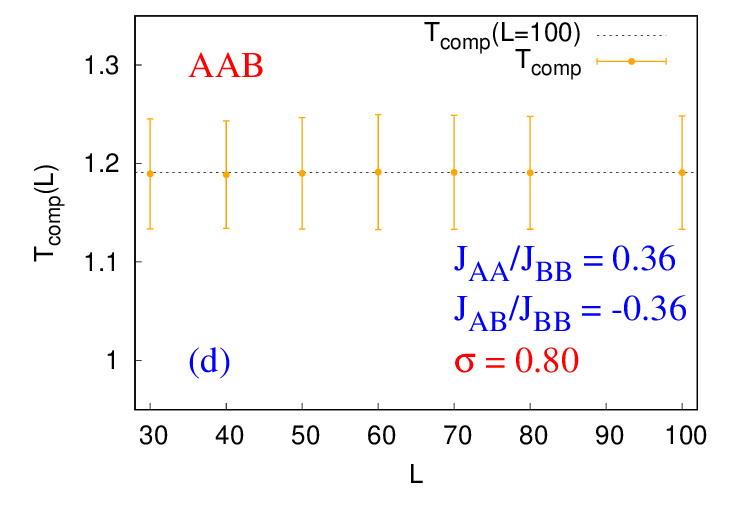}}
			
		\end{tabular}
		\caption{ (Colour Online) Compensation temperatures versus linear system sizes of Ising trilayered square stacking of ABA type (a \& b) and AAB type (c \& d). The reported value of compensation temperature (L = 100) is confined within $1\%$ across all the sizes.}
		\label{fig_tcomp_lsize_field}
	\end{center}
\end{figure*}

We have mentioned that the system size in this study is fixed at $L=100$ in Section \ref{sec_simulation}. The size of the system is a key factor in influencing the thermodynamic response in the context of a trilayered Ising system's critical properties in a field-free environment, according to \cite{Diaz2}. Due to the opposing magnetic moments of the sublayers, compensation happens when the net magnetization of the system is zero. In \cite{Diaz2}, for $L\geq60$, we have observed that the compensation point is immune to the linear system size in a field-free environment. But under the influence of a Gaussian random external field with spatio-temporal variation, we still need to find out whether the compensation point of the steady state magnetisations depends on the linear system size. To address this issue, in this section, a few representative cases are discussed in Figure \ref{fig_tcomp_lsize_field}. For both the configurations, ABA and AAB, the combination of moderate ferromagnetic-moderate antiferromagnetic coupling ratio is shown with $J_{AA}/J_{BB}=0.36$ and $J_{AA}/J_{BB}=-0.36$ and system sizes varied from $L=30$ to $L=100$. From these cases, we see the fluctuations in the values of compensation points have been confined to within $1\%$ across system sizes. For very few other randomly checked combinations of coupling strengths, the same feature is present. So, within the scope of available limited computational resources, using only the values of compensation temperatures for linear system sizes $L = 100$ doesn't compromise much on the accuracy for all the combinations of coupling strengths and standard deviation of the external Gaussian random magnetic field. That is a strong compendium in support of choosing $L=100$, in this study.

%%%%%%%%%%%%%%%%%%%%%%%%%%%%%%%%%%%%%%%%%%%%
%%%%%%%%%%%%%%%%%%%%%%%%%%%%%%%%%%%%%%%%%%%%%
\section{On the transient behaviour}
\label{app_transient}
%%%%%%%%%%%%%%%%%%%%%%%%%%%%%%%%%%%%%%%%%%%%%
In this study, we have magnetic systems which are responding to time-dependent external fields and achieving steady-state. So we need to figure out the time interval the systems usually consume to die out the transient behaviours. A few selected examples are provided here which shows the reason behind the choice of transient time interval $(\Delta T)$ in Section \ref{sec_simulation}. We can roughly estimate that $\Delta T \sim 50$ MCS, and that's why $250$ MCS is consumed to reach the steady state (and record data) for surity.

\begin{figure*}[!htb]
	\begin{center}
		\begin{tabular}{c}
			
			\resizebox{8.5cm}{!}{\includegraphics[angle=0]{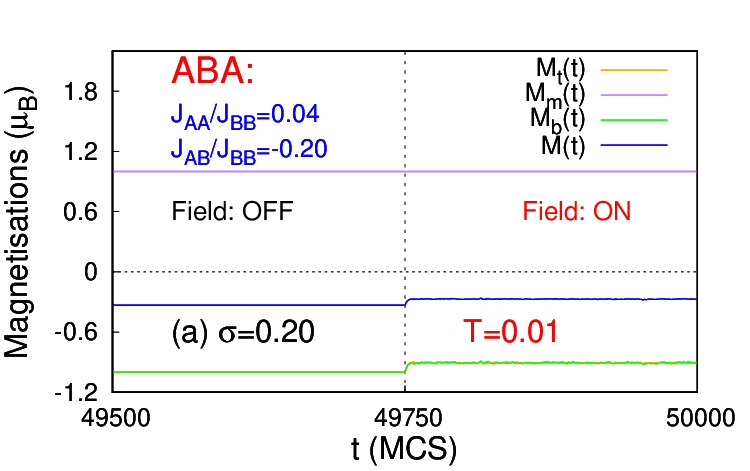}}
			
			\resizebox{8.5cm}{!}{\includegraphics[angle=0]{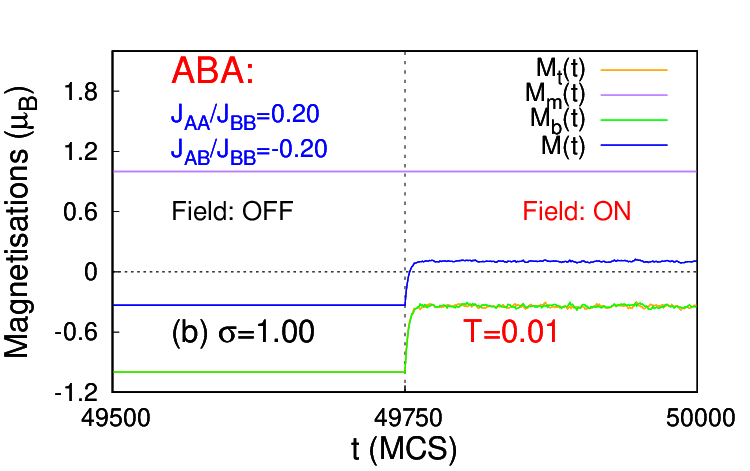}}\\
			
			\resizebox{8.5cm}{!}{\includegraphics[angle=0]{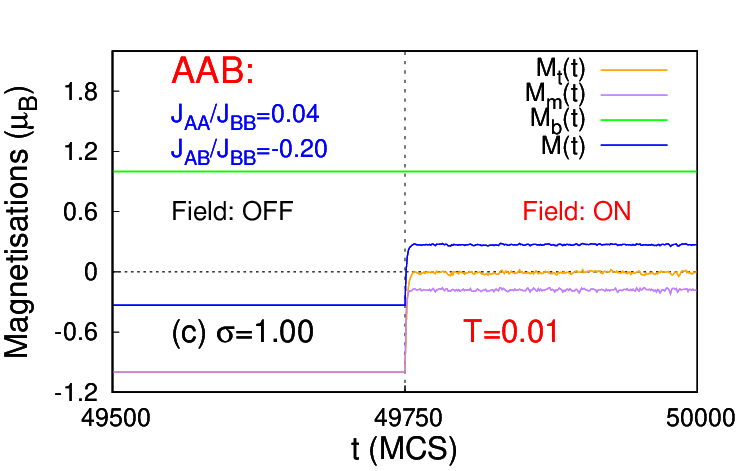}}
			
			\resizebox{8.5cm}{!}{\includegraphics[angle=0]{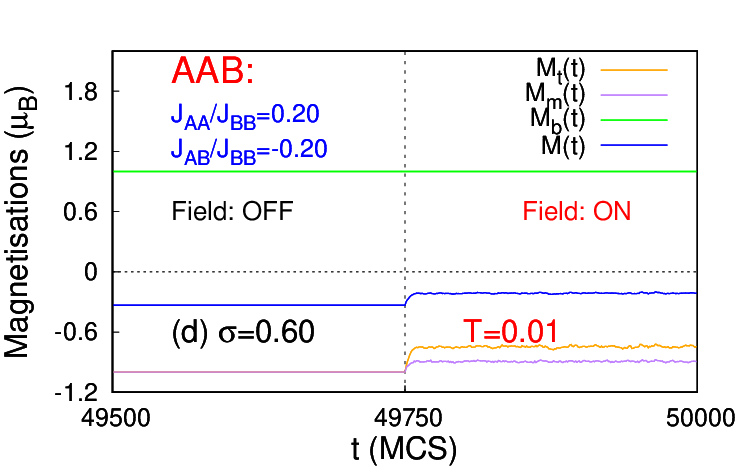}}
			
		\end{tabular}
		\caption{ (Colour Online) Transient behaviour of the trilayered systems: ABA (a \& b) and AAB (c \& d) for a few cases.}
		\label{fig_transient}
	\end{center}
\end{figure*}

%%%%%%%%%%%%%%%%%%%%%%%%%%%%%%%%%%%%%%%%%%%%%
\section{Magnetic behaviour for $\sigma=1.00$}
\label{app_high_sigma}
\begin{figure*}[!htb]
	\begin{center}
		\begin{tabular}{c}			
			\Large{\textbf{ABA: $\sigma=1.00$}}\\
			\\
		
			\phantom{}
			\hspace{0.00cm} $\mathbf{J_{AA}/J_{BB}=0.04}$ \hspace{3.00cm} $\mathbf{J_{AA}/J_{BB}=0.20}$ \hspace{3.00cm} $\mathbf{J_{AA}/J_{BB}=1.00}$ \\
			
			\hspace{0.00cm} $\mathbf{J_{AB}/J_{BB}=-0.20}$ \hspace{2.75cm} $\mathbf{J_{AB}/J_{BB}=-0.20}$ \hspace{2.75cm} $\mathbf{J_{AB}/J_{BB}=-0.52}$ \\
			\\
			
			\resizebox{6.0cm}{!}{\includegraphics[angle=0]{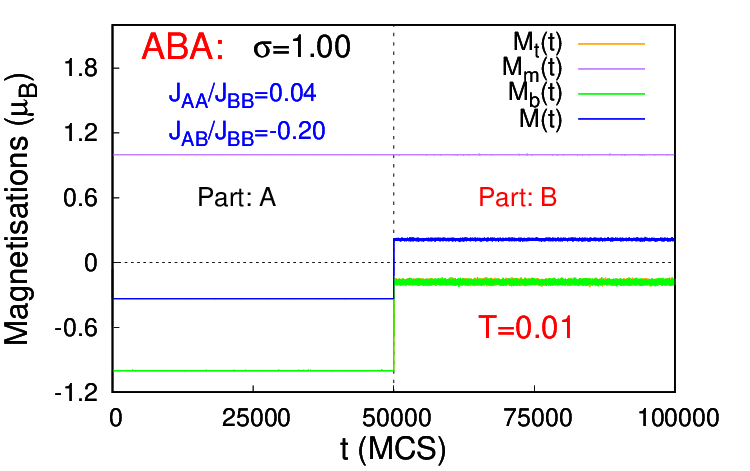}}
			
			\resizebox{6.0cm}{!}{\includegraphics[angle=0]{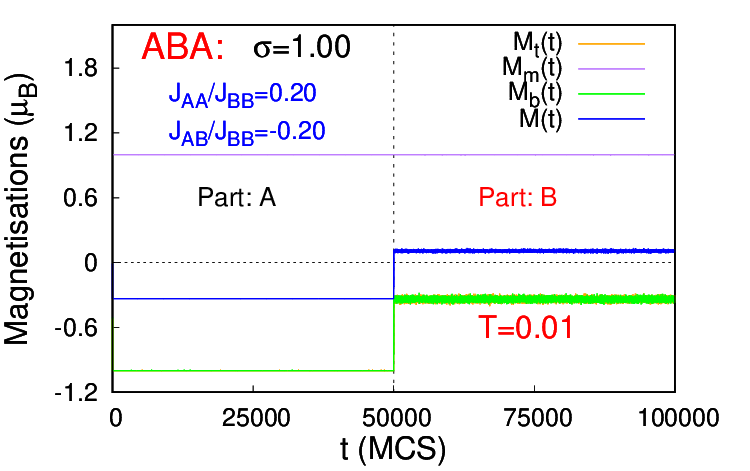}}
			
			\resizebox{6.0cm}{!}{\includegraphics[angle=0]{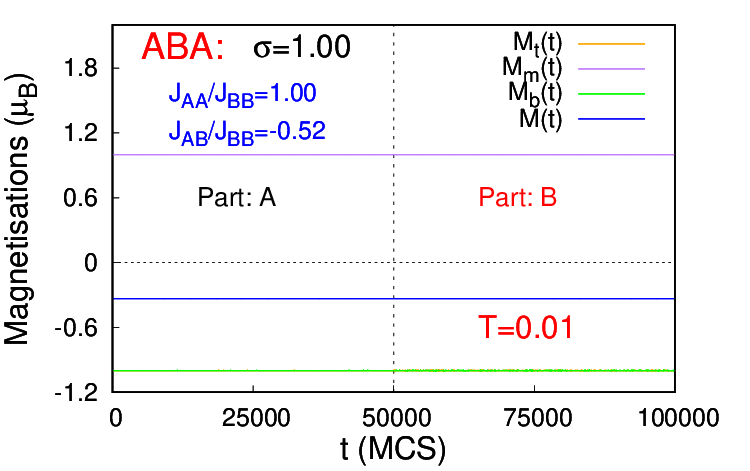}}\\
			
			\\

			\Large{\textbf{AAB: $\sigma=1.00$}}\\
			\\
			
			\phantom{}
			\hspace{0.00cm} $\mathbf{J_{AA}/J_{BB}=0.04}$ \hspace{3.00cm} $\mathbf{J_{AA}/J_{BB}=0.20}$ \hspace{3.00cm} $\mathbf{J_{AA}/J_{BB}=1.00}$ \\
			
			\hspace{0.00cm} $\mathbf{J_{AB}/J_{BB}=-0.20}$ \hspace{2.75cm} $\mathbf{J_{AB}/J_{BB}=-0.20}$ \hspace{2.75cm} $\mathbf{J_{AB}/J_{BB}=-0.52}$ \\
			\\
			
			\resizebox{6.0cm}{!}{\includegraphics[angle=0]{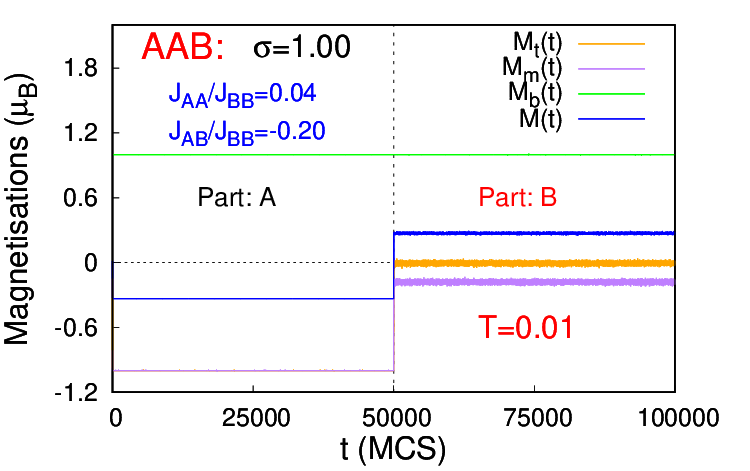}}
			
			\resizebox{6.0cm}{!}{\includegraphics[angle=0]{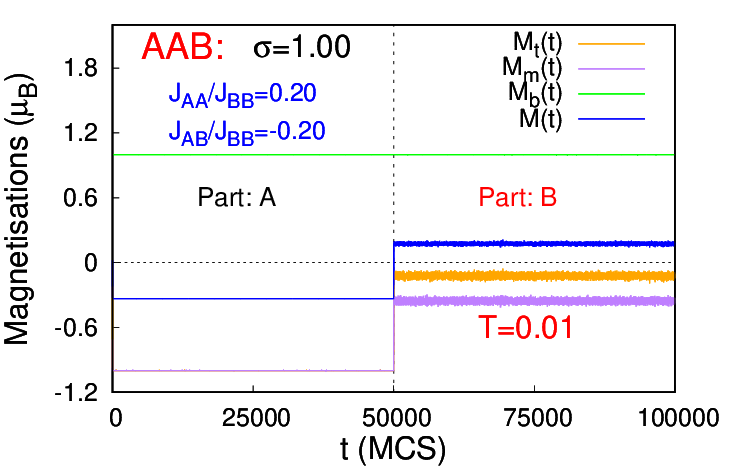}}
			
			\resizebox{6.0cm}{!}{\includegraphics[angle=0]{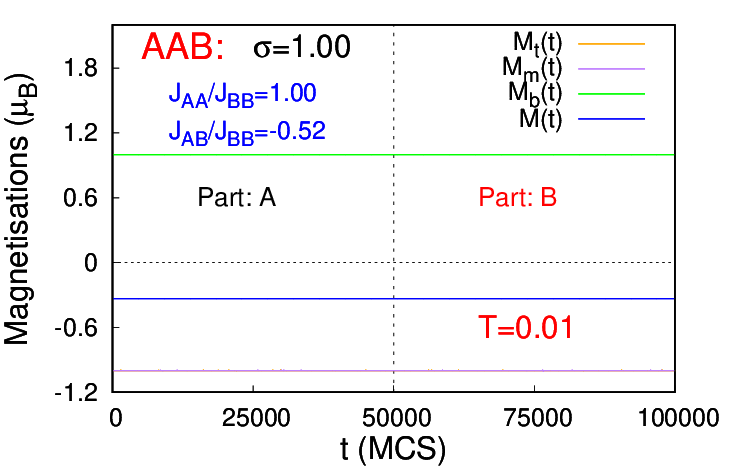}}\\
			
			\\
			
		\end{tabular}
		\caption{ (Colour Online) Plots of Magnetisations for square monolayers (sublattices) and total magnetisation of the bulk versus time in MCS, \textbf{for the \textit{ABA} and \textit{AAB} configurations} with $\sigma=1.00$ . Here, $M_{t}(t)$: Magnetization of the top layer; $M_{m}(t)$: Magnetization of the mid layer; $M_{b}(t)$: Magnetization of the bottom layer are all functions of time,$t$, in units of MCS. In these figures, \textbf{Part: A} describes the equilibrium (Zero-field) and transient (Field: ON) behaviour whereas, \textbf{Part: B} describes the steady state behaviour (Field: ON).}
		\label{fig_max_sigma}
	\end{center}
\end{figure*}

%%%%%%%%%%%%%%%%%%%%%%%%%%%%%%%%%%%%%%%%%%%%%
In this section, a small add-on is provided for better understanding of how magnetic order diminishes when the standard deviation is increased to $\sigma=1.00$ and dimensionless temperature, $T=0.01$. \textbf{In the field-free environment, all the sublayers are perfectly ordered in such a nearly athermal condition}. A few examples are provided in Figure \ref{fig_max_sigma} where the external field affects the magnetic behaviour even in such low temperature. We understand now that when in-plane coupling strengths and corresponding cooperative energies are comparable to the steady state spin-field energies, the corresponding sublayer magnetisations don't deviate much from equilibrium values. But when the spin-field term dominates, the steady state sublayer magnetisation diminishes gradually as we increase the randomness of the external Gaussian random field. If the in-plane coupling strength is very weak, e.g. $J_{AA}/J_{BB}=0.04$ for the AAB configuration in Figure \ref{fig_max_sigma}(B), magnetic ordering in the steady state almost vanishes (steady-state sublayered magnetisation $\approx$ 0).

%%%%%%%%%%%%%%%%%%%%%%%%%%%%%%%%%%%%%%%%%%%%
\vskip 2 cm
\begin{center} {\Large \textbf {References}} \end{center}
%%%%%%%%%%%%%%%%%%%%%%%%%%%%%%%%%%%%%%%%%%%%%

%%%%%%%%%%%%%%%%%%%%%%%%%%%%%%%%%%%%%%%%%%%%%%
\end{document}